\documentstyle[twoside,fleqn,espcrc2,epsfig]{article}

\newcommand{\AmS}{{\protect\the\textfont2
  A\kern-.1667em\lower.5ex\hbox{M}\kern-.125emS}}
\let\q=\quad

\def\vs{{\it vs.}}
\def\etal{{\it et al.}}
\def\etc{{\it etc.}}
\def\eg{{\it e.g.}}
\def\antibar#1{\overline{#1}}
\def\nubar{\antibar{\nu}}

\def\nutau{{\nu_\tau}}
\def\mnutau{m_{\nu_\tau}}

\def\lsim{\stackrel{<}{\scriptstyle\sim}}
\def\gsim{\stackrel{>}{\scriptstyle\sim}}
\def\BR{{\cal B}}

\def\vev#1{\left\langle #1\right\rangle}

\def\lra{\leftrightarrow}

\title{TAU 98 Conference Summary}

\author{Alan J.~Weinstein\address{California Institute of Technology 
        Pasadena, CA 91125}}
       
\begin{document}

\begin{abstract}
I very briefly review the highlights of the fifth workshop
on the physics of the tau lepton and its neutrino.
There has been much progress in many sub-fields,
which I touch upon in this review:
the couplings of the tau to the $Z^0$ and $W^\pm$;
the leptonic branching fractions, lifetime, and tests of universality;
the Lorentz structure of tau decays;
searches for neutrinoless decays;
limits on weak and electromagnetic dipole moments and CP violation;
inclusive semi-hadronic decays, spectral functions,
sum rules, QCD, and applications;
substructure in tau decays to three pseudoscalars;
tau decays to kaons; 
limits on the mass of the tau neutrino;
tau neutrinos from solar, atmospheric, and AGN sources;
accelerator searches for neutrino oscillations;
and prospects for the future.
\end{abstract}

\maketitle

%%%%%%%%%%%%%%%%%%%%%%%%%%%%%%%%%%%%%%%%%%%%%%%%%%%%%%%%%%%%%%%%%%%%%%%%%
\section{INTRODUCTION}
\label{s-intro} 

This bi-annual series of workshops is an outstanding opportunity
to focus on the enormous breadth and depth of fundamental physics
accessible from the study of the production and decay 
of the tau lepton and the tau neutrino.
At each meeting, we have seen tremendous progress in the precision
with which Standard Model physics is measured,
and increasing sensitivity to new physics; and 
this meeting, fifth in the series, is no exception.
Tau physics continues to be wonderfully rich and deep!

The study of the tau contributes to the field of particle physics 
in many ways. I think of the ``sub-fields'' as follows:
\begin{itemize}
\item Precision electroweak physics:
      neutral current ($Z^0$) and charged current ($W^\pm$) couplings,
      Michel parameters, leptonic branching fractions
      and tests of universality. These measurements typically
      have world-average errors better than 1\%.
      They give indirect information on new physics at high mass scales
      ($Z^\prime$, $W_R$, $H^\pm$, \etc).
      The higher the precision, the better chance of seeing
      the effects of new physics, so continued improvement is necessary.
\item Very rare (``upper limit'') physics:
      Direct searches for new physics, such as
      processes forbidden in the Standard Model: 
      lepton-flavor-violating neutrinoless decays, 
      lepton-number-violating decays such as $\tau^-\to\mu^+ X$, 
      and CP-violating effects that can result from
      anomalous weak and electromagnetic dipole moments.
\item Non-perturbative hadronic physics:
      Our inability to reliably predict the properties
      and dynamics of light mesons and baryons
      at intermediate energies is the greatest failing
      of particle physics. Tau decays provide a clean beam
      of intermediate energy light mesons,
      including vectors, axial-vectors, scalars and tensors.
      One can measure mesonic couplings, and tune models
      of resonances and Lorentz structure.
      Topics of current interest are the presence
      and properties of radial excitations such as
      the $a_1^\prime$, and $K^{*\prime}$, 
      mixing, SU(3)$_f$ violation, and isospin decomposition
      of multihadronic final states.
\item Inclusive QCD physics: the total and differential
      inclusive rate for tau decays to hadrons (spectral functions),
      can be used to measure $\alpha_S(s)$,
      non-perturbative quark and gluon condensates,
      and quark masses; and can be used to test QCD sum rules.
      Here, in particular, there is a 
      rich interaction between theory and experiment.
\item Neutrino mass and mixing physics:
      Aside from its fundamental significance, the presence
      of tau neutrino mass and mixing has important implications
      for cosmology and astrophysics $-$
      it is our window on the universe.
\end{itemize}

The study of taus is thus an important tool in many fields,
and continual progress is being made in all of them,
as evidenced in the many contributions to this workshop
which I review in the following sections.

Some of the more exciting future goals in tau physics
were discussed in the first talk of the workshop,
by Martin Perl \cite{ref:perl}.
This was followed by a very comprehensive
overview of the theory of tau physics,
including future prospects for testing the theory,
by J.~Kuhn \cite{ref:kuhn}.
These talks are reviews in and of themselves;
I will focus, in this review, on the subsequent presentations.

%---------------------------------------------
\section{$\mathbf{Z^0}$ COUPLINGS}
\label{s-Z0couplings}

SLD and the four LEP experiments study the reaction
$e^+e^- \to Z^0 \to \tau^+\tau^-$ 
to extract a wealth of information
on rates and asymmetries,
and ultimately, on the neutral current
vector and axial couplings $v_f$ and $a_f$
for each fermion species $f$,
and from these, values for the effective weak mixing angle
$\sin^2\theta_W^f$.
Lepton universality is the statement that
these couplings (and the charged current couplings 
to be discussed later) are the same for the electron,
muon and tau leptons.

The simplest such observable is
the partial width of the $Z^0$ into fermion pairs,
which to lowest order is:
$$ \Gamma_f \equiv \Gamma (Z^0 \to f\bar{f}) =
   \frac{G_F M_Z^3}{6\sqrt{2}\pi} 
   \left( v_f^2 + a_f^2 \right). $$
Equivalently, the ratio of the total
hadronic to leptonic width $R_\ell \equiv \Gamma_{had}/\Gamma_\ell$,
with $\ell=e, \mu, \tau$, can be measured with high precion.

The angular distribution of the outgoing fermions
exhibit a parity-violating forward-backward asymmetry
$A_f^{FB} = \frac{3}{4}{\cal A}_e{\cal A}_f$
which permits the measurement of the asymmetry parameters
$${\cal A}_f \equiv 
   \frac{2 v_f a_f}{v_f^2 + a_f^2} .$$
From these one can extract
the neutral current couplings $v_f$ and $a_f$.

The Standard Model predicts that the
outgoing fermions from $Z^0$ decay are polarized,
with a polarization that depends on the scattering angle
$\cos\theta_f$ with respect to the incoming electron or positron:
$${\cal P}_f(\cos\theta_f) =
- \frac{{\cal A}_f(1+\cos^2\theta_f) + 2{\cal A}_e\cos\theta_f}
       {(1+\cos^2\theta_f) + 2{\cal A}_f{\cal A}_e\cos\theta_f} .$$
If the incoming electron beam is also polarized,
as at the SLC, it further modifies the polarization
of the outgoing fermions.
In the case of the tau ($f = \tau$), this polarization can be 
measured at the statistical level by analyzing the 
decay products of the tau, providing an independent way
to determine ${\cal A}_\tau$ and thus $v_\tau$ and $a_\tau$.
LEP and SLD have used almost all of the major tau decay modes
($e$, $\mu$, $\pi$, $\rho$, $a_1$)
to analyze the tau spin polarization as a function of $\cos\theta_\tau$.
SLD has measured the dependence of these polarizations on beam ($e^-$)
polarization.

\begin{figure}[ht]
\psfig{figure=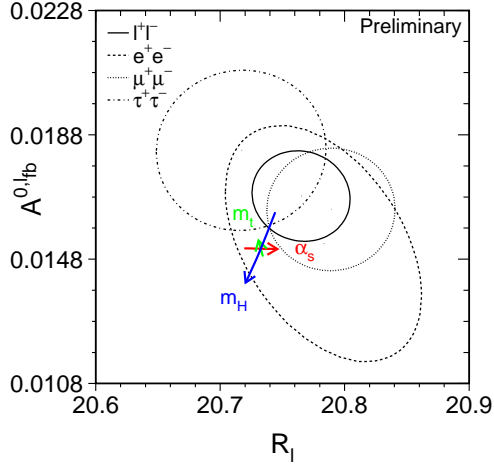,width=2.6in}
\caption[]{LEP-I averages for
$A_{FB}$ versus $R_\ell$ for the three 
lepton species and for the combined result.
The Standard Model prediction is given by the lines \cite{ref:sobie}.}
\label{fig:rl_afb}
\end{figure}

\begin{figure}[ht]
\psfig{figure=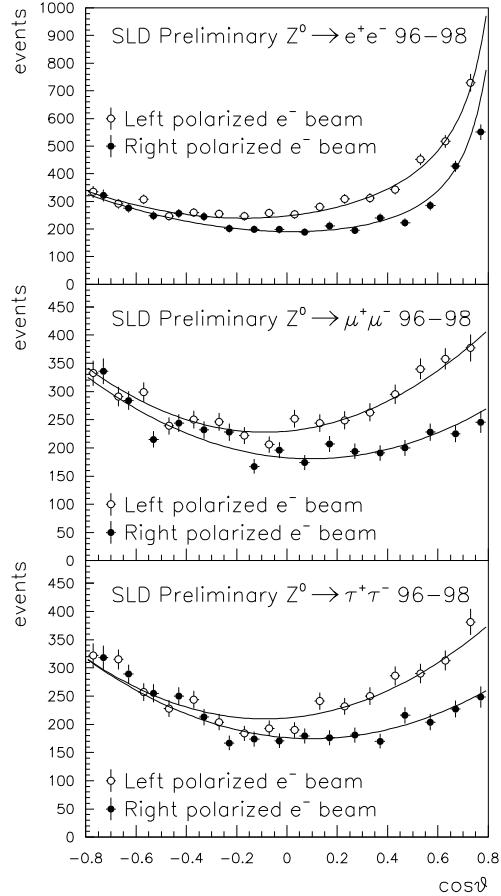,width=2.6in}
\caption[]{Polar angle distributions for leptonic final states,
from SLD with polarized beams \cite{ref:reinertsen}.}
\label{fig:sld}
\end{figure}

SLD and all four LEP experiments measure
all these quantities. 
At TAU 98, the results for the ratios of partial widths $R_\ell$
and forward-backward asymmetries $A_{FB}^\ell$ for $\ell=e, \mu, \tau$
are reviewed in \cite{ref:sobie} and shown in Fig.~\ref{fig:rl_afb}.
The tau polarization measurements at LEP are presented 
in \cite{ref:alemany}.
The beam polarization dependent asymmetries at SLD
are described in \cite{ref:reinertsen}, and shown in Fig.~\ref{fig:sld}.
The procedure for combining the LEP results on ${\cal P}_\tau$
is reviewed in \cite{ref:roney},
in which it is concluded that the results from the four
LEP experiments are consistent but not too consistent.
These are nearly the final results from LEP on this subject.

Note that LEP and SLD have completely consistent results
on the leptonic neutral current couplings;
the addition of measurements of the
heavy quark couplings to their
Standard Model averages for the Weinberg angle 
$\sin^2\theta_W$ pull the LEP average away from that of SLD's
(and that discrepancy is shrinking as LEP and SLD results are updated).

\begin{figure}[ht]
\psfig{figure=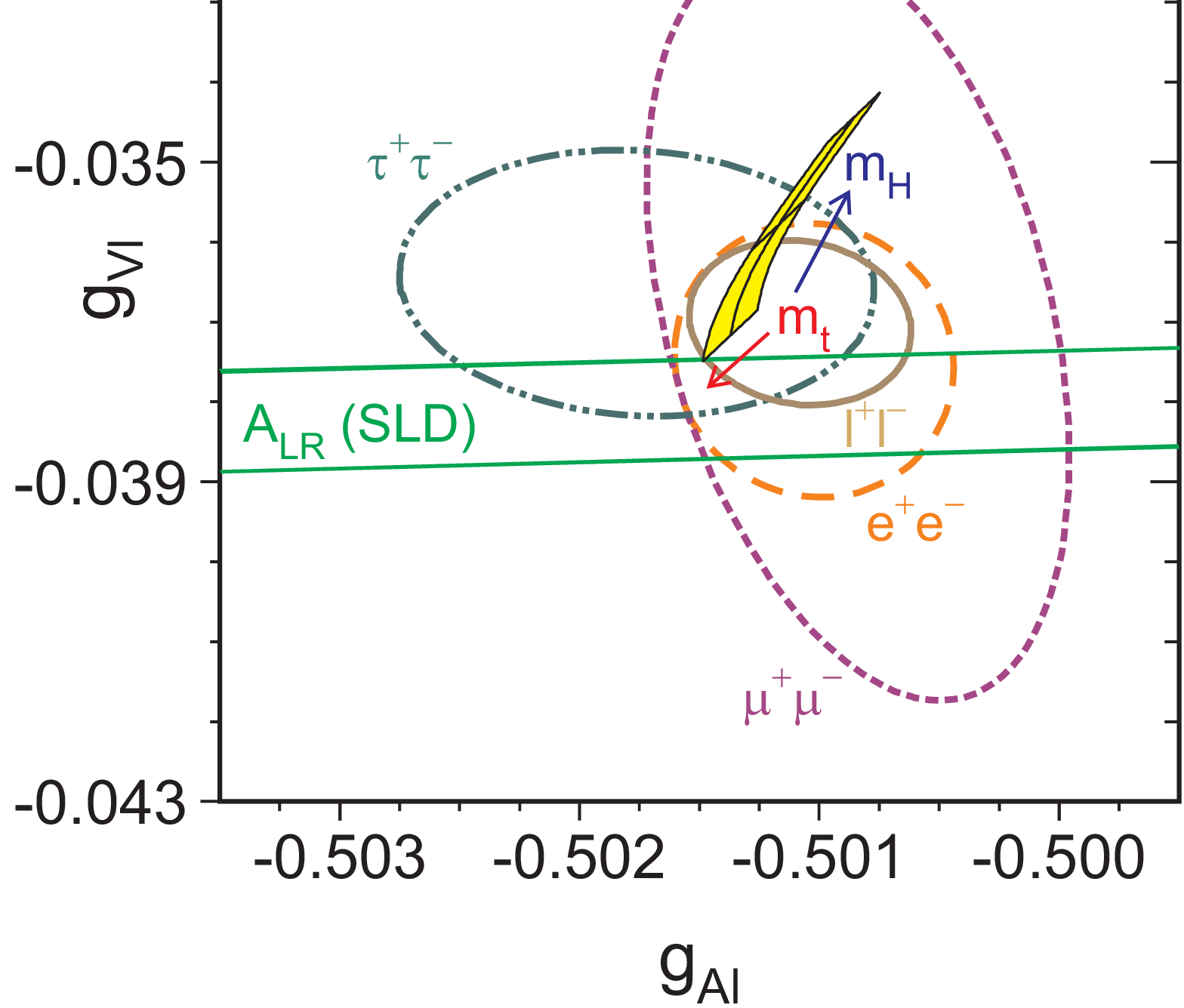,width=2.6in}
\caption[]{Results on the vector and axial couplings of the leptons
to the $Z^0$, from LEP-I \cite{ref:sobie}.}
\label{fig:gvga}
\end{figure}

The partial widths for the three charged leptons
agree with one another (and therefore with lepton universality
in the neutral current) to 3 ppm.
These results, along with those from the $e^+e^-$ and $\mu^+\mu^-$
final states, fit well to the Standard Model predictions
with a rather light Higgs mass:
$m_H < 262$ GeV/c$^2$ at 95\%\ C.L.
The results for the vector and axial couplings
of the three leptons is shown in Fig.~\ref{fig:gvga}.
Note that the tau pair contour is smaller than that for 
mu pairs, because of the added information 
from tau polarization.

%---------------------------------------------
\section{$\mathbf{W\to \tau\nu}$}
\label{s-Wtotau}

\begin{figure}[ht]
\psfig{figure=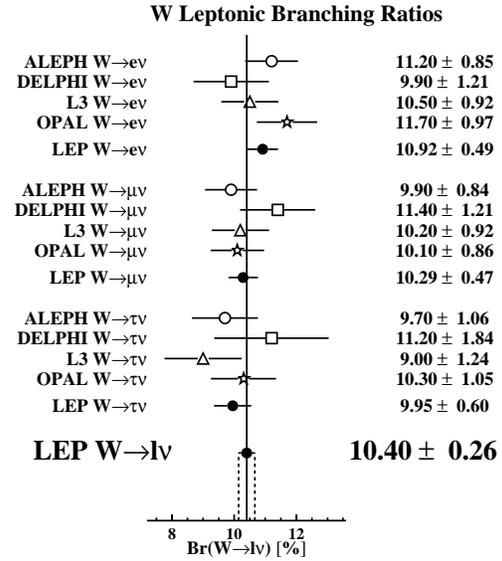,width=2.6in}
\caption[]{Branching fractions for $W\to \ell\nu$ from the 
4 LEP experiments \cite{ref:moulik}.}
\label{fig:wbrlep}
\end{figure}

This TAU workshop is the first to see results from LEP-II,
including new results on
the production of taus from real $W$ decays.
All four LEP experiments identify the leptonic decays of 
$W$ bosons from $e^+e^-\to W^+W^-$ at center-of-mass energies
from 160 to 189 GeV.
The results are consistent between experiments (see Fig.~\ref{fig:wbrlep})
and the averaged branching fractions they obtain 
are summarized in \cite{ref:moulik}:
\begin{eqnarray*}
{\cal B}(W\to e\nu)    &=& 10.92\pm0.49\% \\
{\cal B}(W\to \mu\nu)  &=& 10.29\pm0.47\% \\
{\cal B}(W\to \tau\nu) &=&  9.95\pm0.60\% \\
{\cal B}(W\to \ell\nu) &=& 10.40\pm0.26\% 
\end{eqnarray*}
where the last result assumes universality of the 
charged current couplings.
These results, and results on the measured cross-sections
for $W^+W^-$ as a function of center-of-mass energy, 
are in good agreement with Standard Model predictions.
Lepton universality from real $W$ decays
is tested at the 4\%\ level at LEP:
$g_\mu/g_e = 0.971\pm0.031$,
$g_\tau/g_e = 0.954\pm0.040$.

\begin{figure}[ht]
\psfig{figure=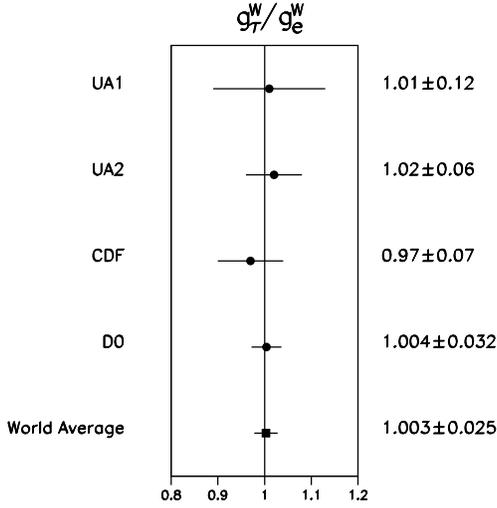,width=2.6in}
\caption[]{Measurements of $g_\tau^W/g_e^W$ 
at hadron colliders \cite{ref:protop}.}
\label{fig:gtauge}
\end{figure}

The CDF and D0 experiments at the Tevatron 
can also identify $W\to \tau\nu$ decays,
and separate them from a very large background
of QCD jets.
The methods and results are summarized in \cite{ref:protop}.
The average of results from UA1, UA2, CDF and D0,
shown in Fig.~\ref{fig:gtauge},
confirm lepton universality in real $W$ decays
at the 2.5\%\ level: 
$g_\tau/g_e = 1.003\pm0.025$.
The LEP and Tevatron
results are to be compared with the ratios of couplings
to virtual $W$ bosons in $\tau$ decays,
summarized in the next section.

\begin{figure}[ht]
\psfig{figure=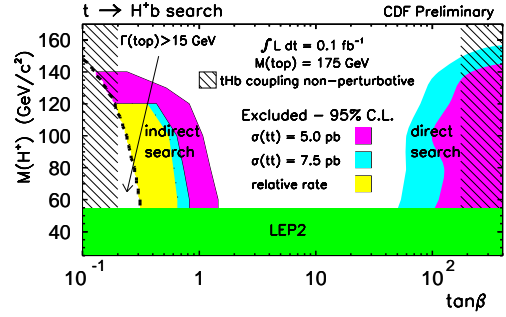,width=2.6in}
\caption[]{Excluded values for the charged Higgs boson mass,
as a function of $\tan\beta$, are shaded \cite{ref:Gallinaro}.}
\label{fig:chiggs}
\end{figure}

CDF has also looked for taus produced from
decays of top quarks,
charged Higgs bosons, leptoquarks, and techni-rhos.
Limits on these processes are reviewed in \cite{ref:Gallinaro}.
For the charged Higgs searches, they are shown
in Fig.~\ref{fig:chiggs}.

%---------------------------------------------
\section{TAU LIFETIME AND
LEPTONIC BRANCHING FRACTIONS}
\label{s-leptonic}

The primary properties of the tau lepton are its mass, spin,
and lifetime. That its spin is 1/2 is well established,
and its mass is well measured \cite{ref:pdg98}:
$m_\tau = 1777.05^{+0.29}_{-0.26}$ GeV/c$^2$. 

The world average tau lifetime has changed considerably
in the last 10 years, 
but recent results have been in good agreement,
converging to a value that is stable and of high precision.

\begin{figure}[ht]
\psfig{figure=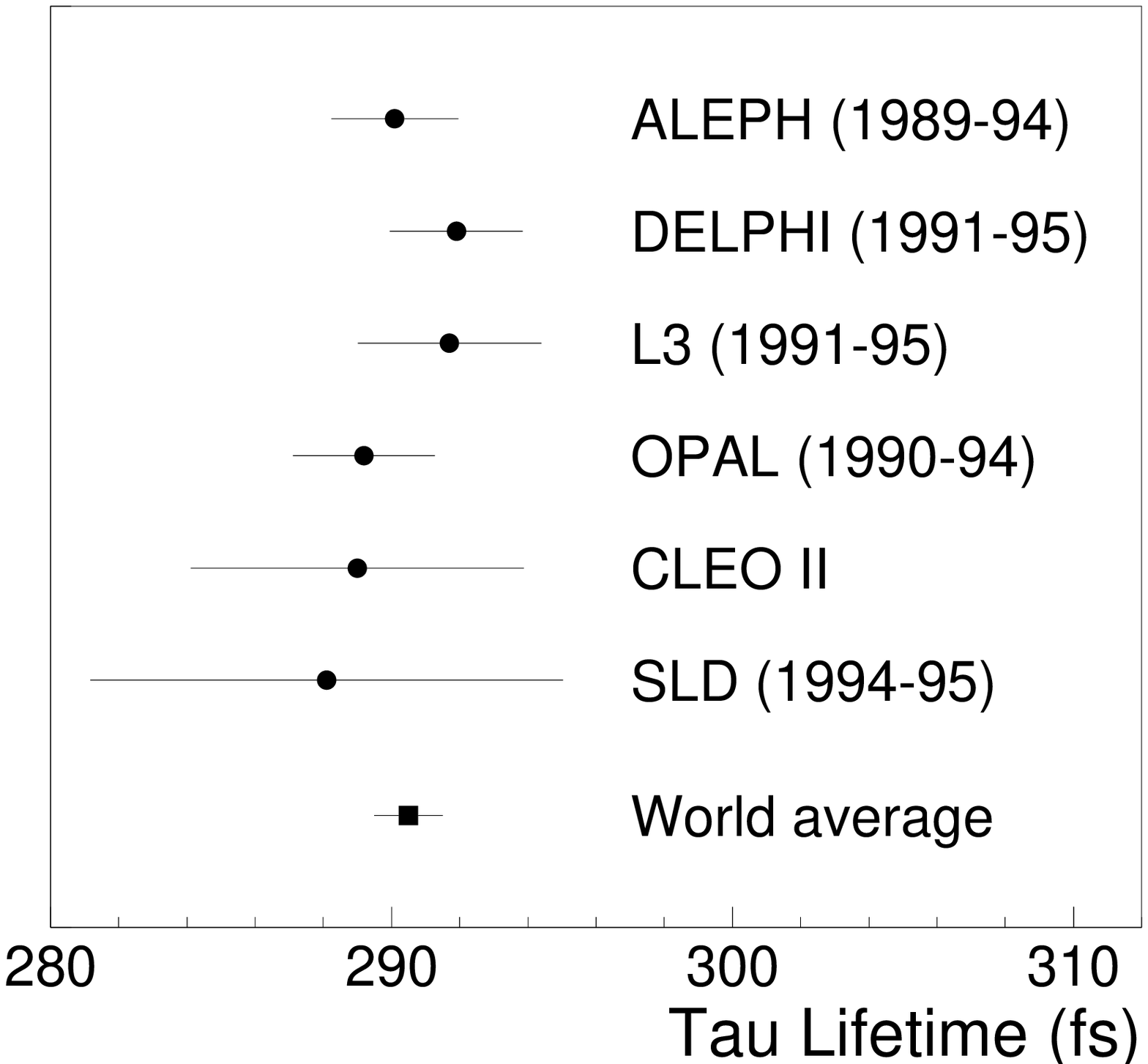,width=2.6in}
\caption[]{Recent measurements of the $\tau$ lifetime \cite{ref:wasserbach}.}
\label{fig:taulife}
\end{figure}

At TAU 98, new measurements were presented from L3 \cite{ref:Colijn} 
and DELPHI.
The new world average tau lifetime represents the work of
6 experiments (shown in Fig.~\ref{fig:taulife}),
each utilizing multiple techniques,
and each with $\lsim 1\%$ precision.
The result is presented in \cite{ref:wasserbach}:
$\tau_\tau = (290.5\pm 1.0)$ fs.

We now turn to the decays of the tau.
The leptonic decays of the tau comprise 35\%\ of the total,
and can be both measured and predicted with high accuracy.
In the Standard Model,
$$\Gamma(\tau\to \ell\nu_\tau\nubar_\ell) = 
\frac{G_{\ell\tau}^2 m_\tau^5}{192\pi^3} 
    f\left(\frac{m_\ell}{m_\tau}\right) (1+\delta).$$
Here, $f$ is a known function of the masses,
$f = 1$ for $e\nu\nu$ and 0.9726 for $\mu\nu\nu$;
$\delta$ is a small correction of $-0.4\%$ 
due to electromagnetic and weak effects, 
and $G_{\ell\tau}$ defines the couplings:
$$G_{\ell\tau} = \frac{g_\ell g_\tau}{4\sqrt{2} m_W^2} = G_F?$$
By comparing the measured leptonic branching fractions
to each other and to the decay rate of the muon to
$e\nu\nu$, we can compare the 
couplings of the leptons
to the weak charged current, $g_e$, $g_\mu$, and $g_\tau$.

At TAU 98, new measurements on leptonic branching fractions
were presented by DELPHI \cite{ref:stugu}, L3 and OPAL \cite{ref:robertson}.
The results are summarized in \cite{ref:stugu}:
\begin{eqnarray*}
{\cal B}(\tau\to e\nu\nubar)    &=& {\cal B}_e = 17.81\pm0.06\% \\
{\cal B}(\tau\to \mu\nu\nubar)  &=& {\cal B}_\mu = 17.36\pm0.06\%.
\end{eqnarray*}
These new world average branching fractions
have an accuracy of 3 ppm,
and are thus beginning to probe the corrections
contained in $\delta$, above.

The resulting ratios of couplings are consistent 
with unity (lepton universality in the charged current couplings)
to 2.5 ppm:
\begin{eqnarray*}
g_\mu/g_e    &=& 1.0014\pm0.0024 \\
g_\tau/g_\mu &=& 1.0002\pm0.0025 \\
g_\tau/g_e   &=& 1.0013\pm0.0025.
\end{eqnarray*}
%The errors are strongly correlated.

Some of the consequences of these precision measurements
are reviewed in \cite{ref:swain}.
The Michel parameter $\eta$ (see next section)
is constrained to be near zero to within 2.2\%.
One can extract a limit on the tau neutrino mass
(which, if non-zero, would cause the function $f$
defined above to depart from its Standard Model value)
of 38 MeV, at 95\%\ C.L.
One can also extract limits on mixing of the tau neutrino
to a 4$^{th}$ generation (very massive) neutrino,
and on anomalous couplings of the tau.

One can even measure fundamental properties of the strong
interaction through precision measurement of these
purely leptonic decays. The total branching fraction
of the tau to hadrons is assumed to be $1-{\cal B}_e -{\cal B}_\mu$.
This inclusive rate for semi-hadronic decays 
can be formulated within QCD:
\begin{eqnarray*}
R_\tau &\equiv& \frac{\BR_h}{\BR_e} =
       \frac{1-\BR_e-\BR_\mu}{\BR_e} = 3.642 \pm 0.019 \\
       &=& 3 (V_{ud}^2 + V_{us}^2) S_{EW} (1+\delta_{pert}+\delta_{NP}),
\end{eqnarray*}
where $V_{ud}$ and $V_{us}$ are Cabibbo-Kobayashi-Maskawa (CKM)
matrix elements,
and $S_{EW}$ is a small and calculable electroweak correction.
The strong interaction between the final state quarks
are described by 
$\delta_{pert}$, the prediction from perturbative QCD
due to radiation of hard gluons, expressible as a 
perturbation expansion in the strong coupling constant 
$\alpha_S(m_\tau^2)$,
and $\delta_{NP}$, which describes non-perturbative 
effects in terms of incalculable expectation values
of quark and gluon operators (condensates).
The value of $\delta_{NP}$ is estimated to be small,
and values for these expectation values can be extracted
from experimental measurements of the spectral functions
in semi-hadronic decays (see section~\ref{s-spectral}).

The extraction of $\alpha_S(m_\tau^2)$ from $R_\tau$
depends on an accurate estimate of $\delta_{NP}$
and on a convergent series for $\delta_{pert}(\alpha_S)$.
Recent improvements in the techniques for calculating
this series (``Contour-improved Perturbation Theory'', CIPT)
and extrapolating to the $Z^0$ mass scale,
are reviewed in ~\cite{ref:Maxwell}.
The resulting values of $\alpha_S$ evaluated 
at the tau mass scale, and then run up to the $Z^0$ mass scale, are:
\begin{eqnarray*}
\alpha_S(m_\tau^2) &=& 0.334\pm 0.010 \\
\alpha_S(m_Z^2   ) &=& 0.120\pm 0.001.
\end{eqnarray*}
We will return to this subject in section \ref{s-spectral}.

%---------------------------------------------
\section{LORENTZ STRUCTURE}
\label{s-lorentz}

The dynamics of the leptonic decays 
$\tau^-\to \ell^-\nu_\ell\nubar_\tau$
are fully determined in the Standard Model, 
where the decay is mediated by the $V-A$ charged current
left-handed $W_L^-$ boson.
In many extensions to the Standard Model,
additional interactions can modify the
Lorentz structure of the couplings, and thus the dynamics.
In particular, there can be 
weak couplings to scalar currents
(such as those mediated by the charged Higgs of the
Minimal Supersymmetric extensions
to the Standard Model, MSSM),
or small deviations from maximal parity violation 
such as those mediated by a right-handed $W_R$ of 
left-right symmetric extensions.

The effective lagrangian for the 4-fermion interaction
between $\tau-\nu_\tau-\ell-\nu_\ell$
can be generalized to include such interactions.
Michel and others in the 1950's
assumed the most general, Lorentz invariant,
local, derivative free, lepton number conserving,
4 fermion point interaction.

Integrating over the two unobserved neutrinos, they
described the differential distribution for the 
daughter charged lepton ($\ell^-$) momentum
relative to the parent lepton ($\mu$ or $\tau$)
spin direction, in terms of the so-called Michel parameters:
\begin{eqnarray*}
\lefteqn{\frac{1}{\Gamma} \frac{d\Gamma}{dxd\cos\theta} 
       = \frac{x^2}{2} \times } \\
&&       \left[ \left( 12(1-x) + \frac{4{ \rho}}{3}(8x-6) 
           + 24{ \eta}{ \frac{m_\ell}{m_\tau}}
                \frac{(1-x)}{x}\right) \right. \\
     &   & \left. \pm { P_\tau }
               { \xi} { \cos\theta} \left( 4(1-x)+ 
        \frac{4}{3}{ \delta}(8x-6) \right)\right] \\
   && \propto  x^2\left[ I(x\vert {\rho , \eta} ) 
          \pm { P_\tau} A( x,{ \theta} \vert
            {\xi ,\delta}) \right],
\end{eqnarray*}
where $\rho$ and $\eta$ are the spectral shape Michel
parameters and $\xi$ and $\delta$ are the spin-dependent 
Michel parameters~\cite{ref:michel};
$x=E_{\ell}/E_{max}$ is the daughter charged lepton energy
scaled to the maximum energy $E_{max} = (m_{\tau}^2 +
m_{\ell}^2)/2m_{\tau}$ in the $\tau$ rest frame;
$\theta$ is the angle between the tau spin direction and the
daughter charged lepton momentum in the $\tau$ rest frame;
and $P_\tau$ is the polarization of the $\tau$.
In the Standard Model
(SM), the Michel Parameters have the values
$\rho=3/4$, $\eta=0$, $\xi = 1$ and $\delta = 3/4$.

There are non-trivial extensions to this approach.
In SUSY models, taus can decay into scalar neutralinos
instead of fermionic neutrinos. 
These presumably are massive,
affecting the phase space for the decay
as well as the Lorentz structure of the dynamics.

In addition, there exists a non-trivial extension
to the Michel formalism that admits anomalous 
interactions with a tensor leptonic current that includes
derivatives; see \cite{ref:seager} for details.
Such interactions will produce distortions
of the daughter charged lepton spectrum
which cannot be described with the Michel parameters.
DELPHI has used both leptonic and semihadronic decays
to measure the tensor coupling $\kappa_\tau^W$,
with the result
$\kappa_\tau^W = -0.029\pm 0.036\pm 0.018$
(consistent with zero).

There are new or updated results on Michel parameter measurements
for TAU 98 from ALEPH, OPAL, DELPHI, and CLEO
\cite{ref:michel}.
The world averages, summarized in \cite{ref:stahl},
also include results from ARGUS, L3, and SLD. 
All results are consistent with the Standard Model,
revealing no evidence for departures from the $V-A$ theory.
These measurements have now reached rather high precision, 
but they are still not competitive with the precision
on Michel parameters obtained from muon decay,
$\mu \to e \nu \nubar$.

\begin{figure}[ht]
\psfig{figure=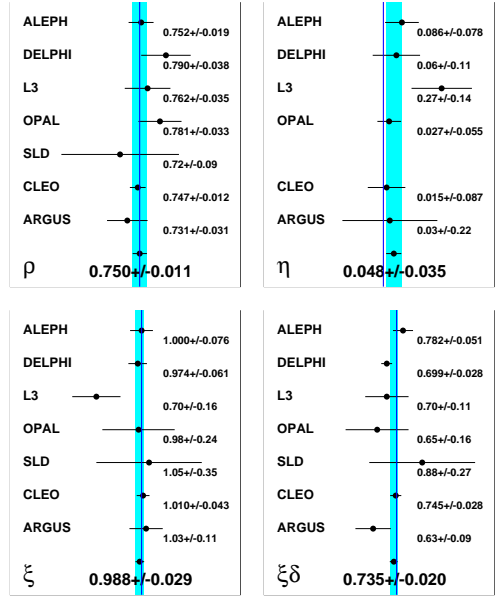,width=2.6in}
\caption[]{New world averages for the Michel parameters in leptonic $
\tau$ decays, assuming $e-\mu$ universality in the couplings \cite{ref:stahl}.}
\label{fig:michwa}
\end{figure}

Results from the two decays $\tau\to e\nu\nu$ and $\tau\to \mu\nu\nu$
can be combined under the assumption of $e-\mu$ universality.
Such an assumption is clearly not called for
when one is searching for new physics that explicitly 
violates lepton universality, such as charged Higgs
interactions, which couple to the fermions 
according to their mass. 
However, such couplings mainly affect the Michel parameter $\eta$,
and it is clear from the Michel formula above
that $\eta$ is very difficult to measure
in $\tau\to e\nu\nu$, since it involves a chirality flip
of the daughter lepton, which is suppressed for the light electron.
Thus, lepton universality is usually invoked to constrain
the $\rho$, $\xi$, and $\xi\delta$ parameters to be the same
for the two decays. Since the measurements of $\rho$ and $\eta$ 
in $\tau\to \mu\nu\nu$ decays are strongly correlated,
the constraint that $\rho_e = \rho_\mu \equiv \rho_{e\mu}$
significantly improves the errors on $\eta_\mu$.
Invoking universality in this sense, 
the world averages for the four Michel parameters \cite{ref:stahl},
shown in Fig.~\ref{fig:michwa}, are:
\begin{eqnarray*}
\rho_{e\mu} &=& 0.7490\pm 0.0082  \q (SM = 3/4) \\
\eta_{e\mu} &=& 0.052 \pm 0.036   \q (SM = 0) \\
\xi_{e\mu}  &=& 0.988 \pm 0.029   \q (SM = 1) \\
(\xi\delta)_{e\mu} &=& 0.734 \pm 0.020 \q (SM = 3/4). 
\end{eqnarray*}

\begin{figure}[ht]
\psfig{figure=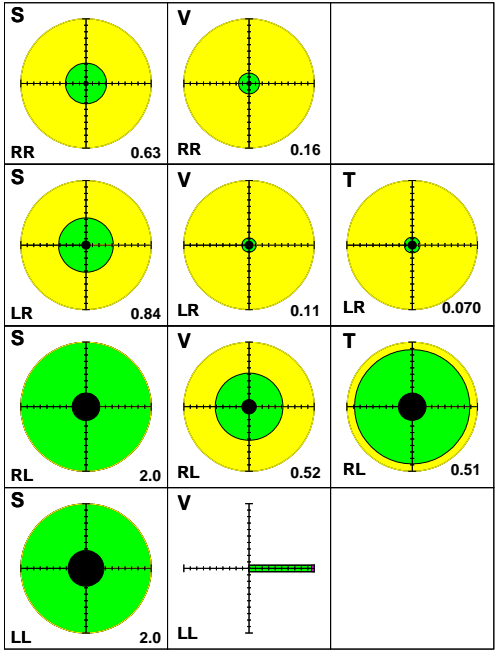,width=2.6in}
\caption[]{Limits on the coupling constants $g^\kappa_{\epsilon\rho}$
in $\tau$ decays, assuming $e-\mu$ universality.
Here, $\kappa = S,V,T$ for scalar, vector, and tensor couplings,
and $\epsilon$ and $\rho$ are the helicities ($L$ or $R$) 
of the $\nu_\tau$ and $\nu_\ell$, respectively.
The black circles are the corresponding
limits from $\mu$ decays \cite{ref:stahl}.}
\label{fig:michcoup}
\end{figure}

A measurement of the spin-dependent Michel parameters
allows one to distinguish the Standard Model $V-A$ interaction
(left-handed $\nu_\tau$) from $V+A$ (right-handed $\nu_\tau$).
The probability that a right-handed (massless) tau neutrino
participates in the decay can be expressed as
$$P^\tau_R = 1/2 \left[1+ 1/9 \left(3 {\xi}
              -16{\xi\delta} \right)\right], $$
and $P^\tau_R = 0$ for the SM $V-A$ interaction.
The Michel parameters measured in all experiments provide
strong constraints on right-handed $(\tau - W - \nu)_R$ couplings,
as shown in Fig.~\ref{fig:michcoup}.
However, they are unable to distinguish left-handed 
$(\tau - W - \nu)_L$ couplings,
for example, between scalar, vector, and tensor currents,
without some additional information, such as 
a measurement of the cross-section
$\sigma(\nu_\tau e^-\to \tau^- \nu_e)$.

In the minimal supersymmetric extension to the 
Standard Model (MSSM), a charged Higgs boson
will contribute to the decay of the tau
(especially for large mixing angle $\tan\beta$),
interfering with the left-handed $W^-$ diagram,
and producing a non-zero value for $\eta$.
Since the world average value of $\eta$ 
(from spectral shapes and the indirect limit from
${\cal B}(\tau\to \mu\nu\nubar)$)
is consistent
with zero, one can limit the mass of a charged Higgs
boson to be \cite{ref:stahl}:
$M(H^\pm) > 2.1 \tan\beta$ (in GeV/c$^2$)
at 95\%\ C.L., which is competitive with direct searches
only for $\tan\beta > 200$.

In left-right symmetric models,
there are two sets of weak charged bosons $W^\pm_1$ and $W^\pm_2$,
which mix to form the observed ``light''
left-handed $W^\pm_L$ and a heavier (hypothetical)
right-handed $W^\pm_R$.
The parameters in these models are
$\alpha = M(W_1)/M(W_2)$ (= 0 in the SM), and
$\zeta = $ mixing angle, $= 0$ in the SM.
The heavy right-handed $W^\pm_R$
will contribute to the decay of the tau,
interfering with the left-handed $W^-$ diagram,
and producing deviations from the Standard Model values
for the Michel parameters $\rho$ and $\xi$.
The limit on $M(W_R)$ is obtained from a 
likelihood analysis \cite{ref:stahl}
which reveals a very weak minimum (less than 1$\sigma$)
at around 250 GeV, so that the 95\%\ C.L.~limit on the mass
of 214 GeV (for a wide range of mixing angles $\zeta$)
is actually slightly worse than it was at TAU 96.
The limit from muon decay Michel parameters is 549 GeV.

It is worth continuing to improve the precision
on the tau Michel parameters, to push the
limits on charged Higgs and right-handed $W$'s,
and perhaps open a window on new physics
at very high mass scales.

%---------------------------------------------
\section{SEARCHES FOR NEW PHYSICS}
\label{s-searches}

One can look directly for physics beyond the Standard Model
in tau decays by searching for decays which violate
lepton flavor (LF) conservation or lepton number (LN) conservation.
These two conservation laws are put into the Standard Model
by hand, and are not known to be the result of some symmetry.
The purported existence of neutrino mixing implies that 
LF is violated at some level,
in the same sense as in the quark sector.
Four family theories,
SUSY, superstrings,
and many other classes of models also predict
LF violation (LFV).
If LF is violated, decays such as $\tau\to\mu\gamma$
become possible; and in general, decays containing
no neutrino daughters are possible (neutrinoless decays).
In theories such as GUTs, leptoquarks, \etc,
a lepton can couple directly to a quark,
producing final states where LN is violated (LNV)
but $B-L$ (baryon number minus lepton number) is conserved.
In most cases, LFV and LNV are accompanied by
violation of lepton universality.

Examples of lepton flavor violating decays
which have been searched for at CLEO, ARGUS, SLC, and LEP include: \\
        $\tau^- \to \ell^-\gamma$, $\ell^-\ell^+\ell^-$ \\
        $Z^0 \to \tau^- e^+$, $\tau^-\mu^+$ \\
        $\tau^- \to \ell^- M^0$, $\ell^- P_1^+ P_2^-$ \\
where the $P$'s are pseudoscalar mesons.
Decays which violate lepton number but conserve $B-L$ include
$\tau^- \to \bar{p} X^0$,
where $X^0$ is some neutral, bosonic hadronic system.
Decays which violate lepton number and $B-L$ include:
        $\tau^- \to \ell^+ P_1^- P_2^-$.

A broad class of R-parity violating
SUSY models which predict LFV or LNV
through the exchange of lepton superpartners
were reviewed at this conference, in \cite{ref:kong}.
A different class of models containing heavy
singlet neutrinos, which produce LFV,
is discussed in \cite{ref:ilakovac}.
In many cases, branching fractions for 
neutrinoless tau decay can be as high as $10^{-7}$
while maintaining consistency with existing data from
muon and tau decays.

CLEO has searched for 
40 different neutrinoless decay modes \cite{ref:stroynowski},
and has set upper limits on the branching fractions
of $\lsim$ few $\times 10^{-6}$.
A handful of modes that have not yet been studied by CLEO,
including those containing anti-protons,
have been searched for by the Mark II and ARGUS experiments,
with branching fraction upper limits 
in the $10^{-4} - 10^{-3}$ range.

Thus, the present limits are approaching levels
where some model parameter space can be excluded.
B-Factories will push below $10^{-7}$; 
it may be that the most important results coming from
this new generation of ``rare $\tau$ decay experiments''
will be the observation of lepton flavor or lepton number violation.

%---------------------------------------------
\section{CP VIOLATION IN TAU DECAYS}
\label{s-cpv}

The minimal Standard Model contains no mechanism for
CP violation in the lepton sector.
Three-family neutrino mixing can produce
(presumably extremely small) violations of CP
in analogy with the CKM quark sector.

CP violation in tau production can occur 
if the tau has a non-zero electric dipole moment or
weak electric dipole moment,
implying that the tau is not a fundamental 
(point-like) object. 
Although other studies of taus (such as production cross section,
Michel parameters, \etc) are sensitive to tau substructure,
``null'' experiments such as the search for CP violating
effects of such substructure can be exquisitely sensitive.
I discuss searches for dipole moments in the next section.

CP violation in tau decays can occur, for example,
if a charged Higgs with complex couplings
(which change sign under CP)
interferes with the dominant $W$-emission process:
$$|A(\tau^-\to W^-\nu_\tau) + g e^{i\theta} A (\tau^-\to H^- \nu_\tau)|^2 .$$
If the dominant process produces a phase shift
(for example, due to the $W\to \rho$, $a_1$, or $K^*$ resonance),
the interference will be of opposite sign
for the $\tau^+$ and $\tau^-$, 
producing a measurable CP violation.
The effect is proportional to isospin-violation
for decays such as $\tau\to \pi\pi\nu_\tau$ $3\pi\nu_\tau$;
and SU(3)$_f$ violation for decays such as 
$\tau\to K\pi\nu_\tau$, $K\pi\pi\nu_\tau$.
The various signals for CP violation in tau decays
are reviewed in \cite{ref:tsai}.

CLEO has performed the first direct search for CP violation
in tau decays \cite{ref:kass},
using the decay mode $\tau\to K\pi\nu_\tau$.
The decay is mediated by the usual p-wave vector exchange,
with a strong interaction phase shift provided by the $K^*$ resonance.
CP violation occurs if there is interference with an
s-wave scalar exchange, with a complex weak phase $\theta_{CP}$
and a different strong interaction phase. 
The interference term is CP odd, so CLEO searches for
an asymmetry in a CP-odd angular observable
between $\tau^+$ and $\tau^-$.
They see no evidence for CP violation,
and set a limit on the imaginary part of the complex coupling
of the tau to the charged Higgs ($g$, in units of $G_F/2\sqrt{2}$):
$g\sin\theta_{CP} < 1.7$.

Tests of CP violation in tau decays can also be
made using $2\pi\nu_\tau$ and $3\pi\nu_\tau$, 
and we can look forward to results from such analyses in the future.

%---------------------------------------------
\section{DIPOLE MOMENTS}
\label{s-dipole}

In the Standard Model, the tau couples to the photon and to the $Z^0$
via a minimal prescription, with a single coupling constant
and a purely vector coupling to the photon
or purely $v_f V + a_f A$ coupling to the $Z^0$.
More generally, the tau can couple to the neutral currents
with $q^2$ dependent vector or tensor couplings.
The most general Lorentz-invariant form of the coupling
of the tau to a photon of 4-momentum $q_\mu$
is obtained by replacing the usual $\gamma^\mu$ with
\begin{eqnarray*}
\Gamma^\mu &=& F_1(q^2)\gamma^\mu + \\
           &&  F_2(q^2)\sigma^{\mu\nu}q_\nu
             - F_3(q^2)\sigma^{\mu\nu}\gamma_5 q_\nu .
\end{eqnarray*}
At $q^2 = 0$, we interpret $F_1(0) = q_\tau$ as the electric charge
of the tau, $F_2(0) = a_\tau = (g_\tau-2)/2$ as the 
anomalous magnetic moment, and
$F_3(0) = d_\tau/q_\tau$, where $d_\tau$ is the 
electric dipole moment of the tau.

In the Standard Model, $a_\tau$ is non-zero due to
radiative corrections, and has the value 
$a_\tau \approx \alpha/2\pi \approx 0.001177$.
The electric dipole moment $d_\tau$ is zero
for pointlike fermions; a non-zero value
would violate $P$, $T$, and $CP$.

Analogous definitions can be made for the
{\it weak} coupling form factors
$F^w_1$, $F^w_2$, and $F^w_3$,
and for the weak static dipole moments $a^w_\tau$ and $d^w_\tau$.
In the Standard Model, the weak dipole moments
are expected to be small (the weak electric dipole moment
is tiny but non-zero due to CP violation in the CKM matrix):
$a_\tau^W = -(2.1 + 0.6i)\times 10^{-6}$, and
$d_\tau^W \approx 3\times 10^{-37}$ e$\cdot$cm.
Some extensions to the Standard Model predict
vastly enhanced values for these moments:
in the MSSM, $a_\tau^W$ can be as large as $10^{-5}$
and $d_\tau^W$ as large as a few $\times 10^{-20}$.
In composite models, these dipole moments
can be larger still.
The smallness of the Standard Model expectations
leaves a large window for discovery of
non-Standard Model couplings.

At the peak of the $Z^0$, the reactions
$Z^0\to \tau^+\tau^-$ are sensitive to the
weak dipole moments, while the electromagnetic dipole moments
can be measured at center of mass energies far below the $Z^0$
(as at CLEO or BES), and/or through the study of 
final state photon radiation in
$e^+e^-\to(\gamma^*, Z^0)\to \tau^+\tau^-\gamma$.

Extensive new or updated
results on searches for weak and electromagnetic
dipole moments were presented at TAU 98.
In all cases, however, no evidence for non-zero
dipole moments or CP-violating couplings were seen,
and upper limits on the dipole moments were
several orders of magnitude larger than Standard Model expectations,
and is even far from putting meaningful limits
on extensions to the Standard Model.
There is much room for improvement in both technique
and statistical power, and, as in any search for
very rare or forbidden phenomena, 
the potential payoffs justify the effort.

An anomalously large weak magnetic dipole moment
will produce a transverse spin polarization 
of taus from $Z^0$ decay, leading to (CP-conserving)
azimuthal asymmetries in the subsequent tau decays.
The L3 experiment searched for these asymmetries
using the decay modes $\tau\to\pi\nu_\tau$ and 
$\tau\to\rho\nu_\tau$, and observed none \cite{ref:vidal}.
They measure values for the real and imaginary parts of 
the weak magnetic dipole moment
which are consistent with zero, and set upper limits
(at 95\%\ C.L.):
\begin{eqnarray*}
|Re(a_\tau^w)| &<& 4.5 \times 10^{-3} \\
|Im(a_\tau^w)| &<& 9.9 \times 10^{-3}.
\end{eqnarray*}
They measure, for the real part of the weak electric
dipole moment, a value which is consistent with zero
within a few $\times 10^{-17}$ e$\cdot$cm.

The ALEPH, DELPHI, and OPAL experiments search for
CP violating processes induced by a 
non-zero weak electric dipole moment,
by forming CP-odd observables from the 4-vectors
of the incoming beam and outgoing taus, and the outgoing tau spin vectors.
These ``optimal observables'' \cite{ref:zalite}
pick out the CP-odd terms in the cross section for production
and decay of the $\tau^+\tau^-$ system;
schematically, the optimal observables are defined by:
$$d\sigma \propto |M_{SM} + d^w_\tau M_{CP}|^2,$$
$$O^{Re} = Re(M_{CP})/M_{SM}, \, O^{Im} = Im(M_{CP})/M_{SM}.$$
These observables are essentially CP-odd triple products,
and they require that the spin vectors of the taus be determined
(at least, on a statistical level).
Most or all of the decay modes of the tau
$(\ell,\pi,\rho,a_1)$ are used to spin-analyze the taus.

ALEPH, DELPHI, and OPAL measure the expectation values
$<O^{Re}>$ and $<O^{Im}>$ of these observables,
separately for each tau-pair decay topology.
From these, they extract measurements of the 
real and imaginary parts of $d_\tau^w$.
The combined limits (at 95\%\ C.L.) \cite{ref:zalite} are:
\begin{eqnarray*}
|Re(d_\tau^w)| &<& 3.0 \times 10^{-18} \q\mbox{e$\cdot$cm} \\
|Im(d_\tau^w)| &<& 9.2 \times 10^{-18} \q\mbox{e$\cdot$cm} .
\end{eqnarray*}
They are consistent with the Standard Model,
and there is no evidence for CP violation.

SLD makes use of the electron beam polarization 
to enhance its sensitivity to Im($d_\tau^W$).
Rather than measure angular asymmetries or expectations
values of CP-odd observables,
they do a full unbinned likelihood fit to the observed event
(integrating over unseen neutrinos)
using tau decays to leptons, $\pi$, and $\rho$,
in order to extract limits on the real and imaginary parts 
of both $a_\tau^w$ and $d_\tau^w$.
They obtain preliminary results \cite{ref:barklow} 
which are again consistent with zero, but which have
the best sensitivity to $Im(a_\tau^w)$ and $Im(d_\tau^w)$:
\begin{eqnarray*}
|Re(d_\tau^w)| &=& (18.3\pm7.8) \times 10^{-18} \q\mbox{e$\cdot$cm}; \\
|Im(d_\tau^w)| &=& (-6.6\pm4.0) \times 10^{-18} \q\mbox{e$\cdot$cm}; \\
|Re(a_\tau^w)| &=& (0.7\pm1.2) \times 10^{-3}; \\
|Im(a_\tau^w)| &=& (-0.5\pm0.6) \times 10^{-3}.
\end{eqnarray*}

%---------------------------------------------
\subsection{EM dipole moments}
\label{ss-emdipole}

The anomalous couplings to photons can be probed,
even on the peak of the $Z^0$, by searching for
anomalous final state photon radiation in
$e^+e^- \to \tau^+\tau^- \gamma$.

The L3 experiment 
studies the distribution of photons
in such events, as a function of the photon energy,
its angle with respect to the nearest reconstructed tau,
and its angle with respect to the beam.
In this way, they can distinguish anomalous 
final state radiation from initial state radiation,
photons from $\pi^0$ decays, and other backgrounds.
The effect on the photon distribution
due to anomalous electric couplings is very similar
to that of anomalous magnetic couplings,
so they make no attempt to extract values for
$a_\tau$ and $d_\tau$ simultaneously, but instead
measure one while assuming the other
takes on its Standard Model value.

They see no anomalous photon production \cite{ref:taylor},
and set the limits (at 95\%\ C.L.) 
$-0.052 < a_\tau < 0.058$ and
$|d_\tau| < 3.1\times 10^{-16}$ e$\cdot$cm.

These results should be compared with those for the muon:
\begin{eqnarray*}
a_\mu^{theory} &=& 0.00116591596(67) \\ 
a_\mu^{expt}   &=& 0.00116592350(780) \\ 
d_\mu^{expt}   &=& (3.7\pm3.4) \times 10^{-19} \q\mbox{e$\cdot$cm}.
\end{eqnarray*}
Theoretical progress on the evaluation of $a_\mu$ in the
Standard Model is reviewed in \cite{ref:Czarnecki},
including the important contribution
from tau decays (see section \ref{ss-gminus2} below).
Progress in the experimental measurement at Brookhaven
is described in \cite{ref:Grosse}.
Clearly, there is much room for improvement of the
measurements of the anomalous moments of the tau.

%---------------------------------------------
\section{SPECTRAL FUNCTIONS}
\label{s-spectral}

The semi-hadronic decays of the tau are dominated by
low-mass, low-multiplicity hadronic systems:
$n\pi$, $n\le 6$; $Kn\pi$, $K\bar{K}$, $K\bar{K}\pi$, $\eta\pi\pi$.
These final states are dominated by resonances
($\rho$, $a_1$, $\rho^\prime$, $K^*$, $K_1$, \etc).
The rates for these decays, taken individually,
cannot be calculated from fundamental theory (QCD),
so one has to rely on models, and on
extrapolations from the chiral limit using
chiral perturbation theory.

However, appropriate sums of final states
with the same quantum numbers can be made,
and these semi-inclusive measures of the 
semi-hadronic decay width of the tau
can be analyzed using perturbative QCD.
In particular, one can define spectral functions
$v_J$, $a_J$, $v_J^S$ and $a_J^S$, as follows:
\begin{eqnarray*}
\lefteqn{
\frac{d\Gamma}{dq^2} (\tau \to \hbox{hadrons} + \nutau) =
  {{G_F^2}\over{32 \pi^2 m_\tau^3}} (m_\tau^2-q^2)^2} \\
   &  \times & \left\{ |V_{ud}|^2 \left[ 
     (m_\tau^2+2 q^2) \left(v_1(q^2)+a_1(q^2)\right) \right. \right. \\
   & & \left. + m_\tau^2 \left(v_0(q^2)+a_0(q^2)\right) \right] \\
   & &  +      |V_{us}|^2 \left[ 
     (m_\tau^2+2 q^2) \left(v_1^S(q^2)+a_1^S(q^2)\right) \right. \\
   & &  \left.\left. + m_\tau^2 \left(v_0^S(q^2)+a_0^S(q^2)\right) \right] 
        \right\} .
\end{eqnarray*}
The spectral functions $v$ and $a$ represent the
contributions of the vector and axial-vector hadronic currents
coupling to the $W$.
The subscripts on these functions denote the spin $J$ of the hadronic system, 
and the superscript $S$ denotes states with net strangeness.

The hadronization information contained in the spectral functions
falls in the low-energy domain of strong interaction
dynamics, and it cannot be calculated in QCD.
Nonetheless, many useful relations between the spectral functions 
can be derived.
For example, in the limit of exact $SU(3)_L\times SU(3)_R$ symmetry,
we have $v_1(q^2) = a_1(q^2) = v_1^S(q^2) = a_1^S(q^2)$
and $v_0(q^2) = a_0(q^2) = v_0^S(q^2) = a_0^S(q^2) = 0$.
Relations amongst the spectral functions
depend on assumptions about how the $SU(3)$ symmetry is broken.
The Conserved Vector Current (CVC) hypothesis
requires that $v_0(q^2) = 0$, and that
$v_1(q^2)$ can be related to the total cross-section for
$e^+e^-$ annihilations into hadrons.
Several sum rules relate integrals of these spectral functions,
as described below.

From arguments of parity and isospin,
the final states containing an even number of pions
arise from the vector spectral function,
and those with an odd number of pions
arise from the axial-vector spectral functions $a_1$,
or in the case of a single pion, $a_0$.
Final states containing one or more kaons
can contribute to both types of spectral functions,
since $SU(3)_f$ is violated
(and because of the chiral anomaly).
Experimentally, one can determine 
whether a final state contributes to $v$ or $a$
through a careful analysis of its dynamics.

The spectral functions can be measured experimentally
by adding up the differential distributions
from all the exclusive final states that
contribute to $v$ or $a$, 
in a ``quasi''-inclusive analysis:
$$ v_1  =  \frac{\BR_{v}}{\BR_e}
   \frac{1}{N_{v}}
   \frac{dN_{v}}{dq^2} 
   \frac{M_\tau^8}{\left(m_\tau^2-q^2\right)^2
                   \left(m_\tau^2+2q^2\right)} ,
$$
and similarly for $a$.

For $v(q^2)$, the result is dominated by the $2\pi$ and $4\pi$
final states, with small contributions from $6\pi$, $K\bar{K}$, and others.
For $a(q^2)$, the result is dominated by the $3\pi$ and $5\pi$
final states, with small contributions from $K\bar{K}\pi$ and others.
The $\pi\nu$ and $K\nu$ final states are delta functions
and must be handled separately.

\begin{figure}[!ht]
\psfig{figure=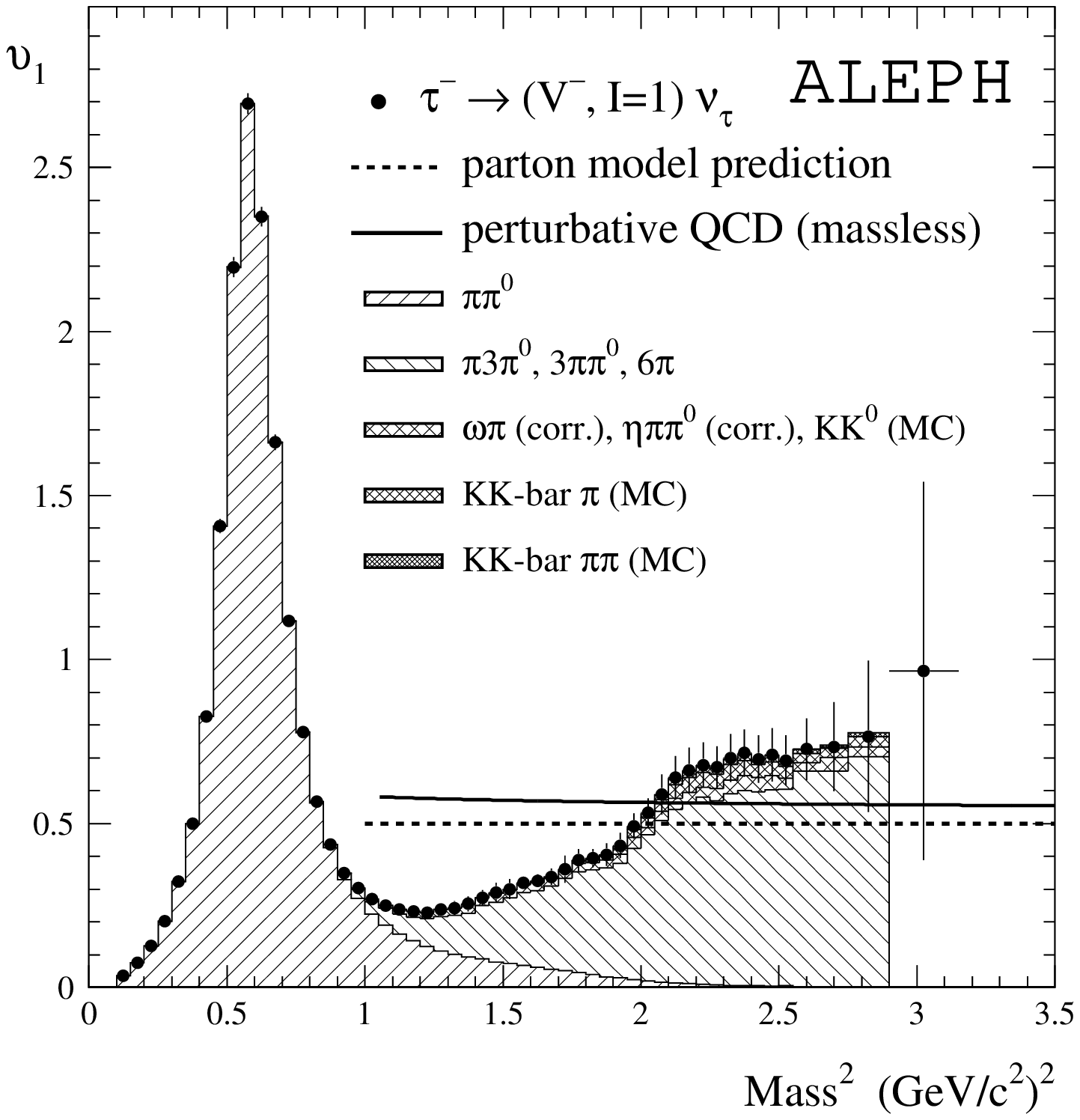,width=2.6in}
\psfig{figure=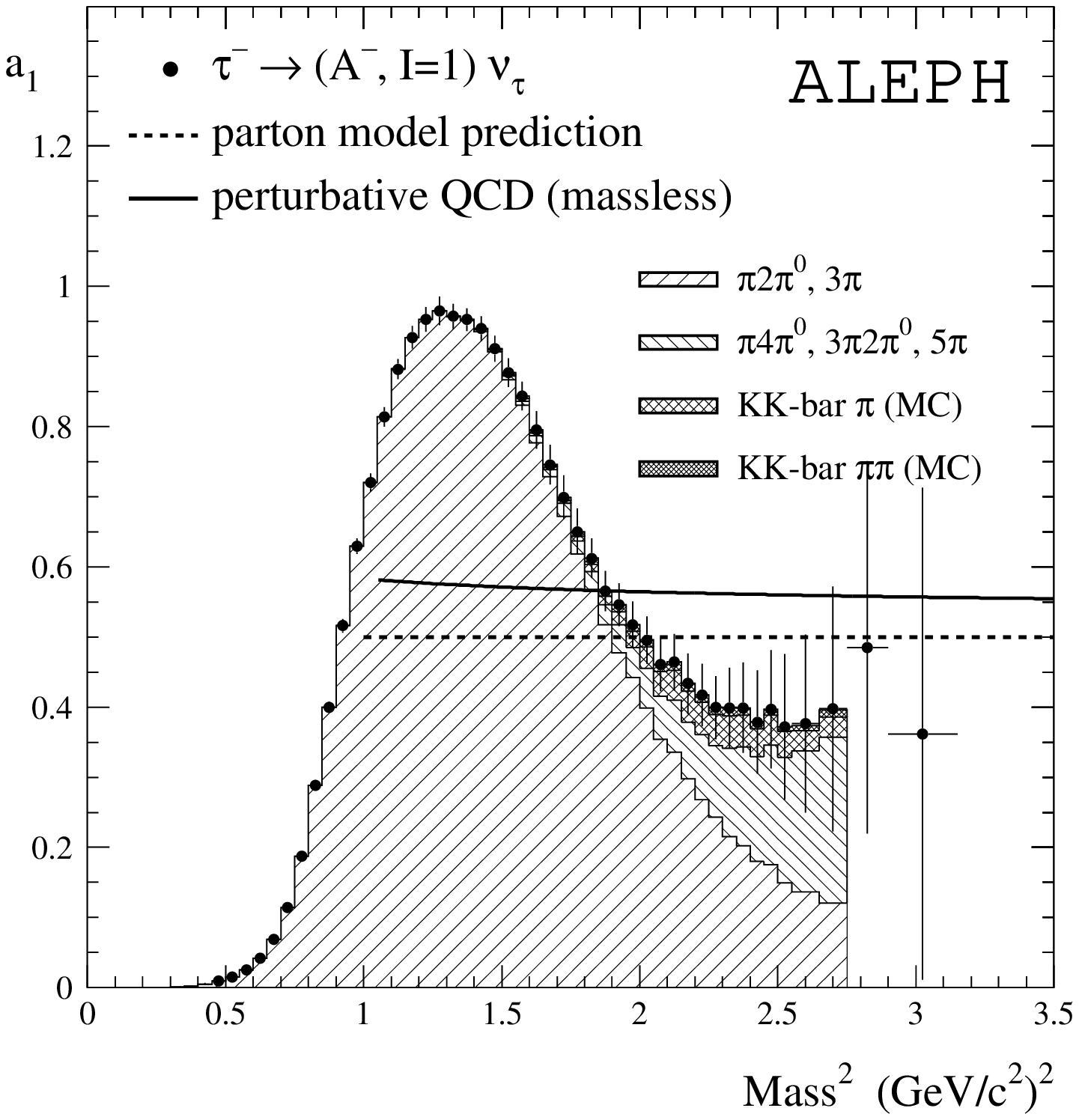,width=2.6in}
\caption[]{Total vector and axial-vector spectral functions from ALEPH.
The contributions from the exclusive channels, from data and MC,
are indicated \cite{ref:Hoecker}.}
\label{fig:spec_aleph}
\end{figure}

OPAL and ALEPH have presented new or updated measurements 
of these non-strange spectral functions \cite{ref:Menke,ref:Hoecker}.
In both cases,
the small contributions mentioned above were obtained
from Monte Carlo estimates, not the data.
the ALEPH results are shown in Fig.~\ref{fig:spec_aleph}.

These spectral functions can be used to study many
aspects of QCD, as described in the following subsections.

%---------------------------------------------
\subsection{Moments of the Spectral functions}
\label{ss-moments}

Although the spectral functions themselves cannot be predicted
in QCD, the moments $R_{kl}$ of those functions:
$$ R_{kl}^{v/a} = \int^{m_\tau^2}_0 ds \left(1 - \frac{s}{m_\tau^2}\right)^k
                                 \left(\frac{s}{m_\tau^2}\right)^l
   \frac{1}{N_{v/a}}
   \frac{dN_{v/a}}{ds},
$$
with $k = 1$, $l = 0 \cdots 3$ {\it are} calculable.
In direct analogy with $R_\tau$ (section \ref{s-leptonic}),
the moments (for non-strange final states) can be expressed as: 
$$R_{kl}^{v/a} = \frac{3}{2} V_{ud}^2 S_{EW} (1+\delta_{pert}
       +\delta_{mass}^{v/a}+\delta_{NP}^{v/a}),$$
where $V_{ud}$ is the CKM matrix element,
and $S_{EW}$ is a small and calculable electroweak correction.
$\delta_{pert}$ is a calculable polynomial
in the strong coupling constant $\alpha_S(m_\tau^2)$,
$\delta_{mass}^{v/a}$ is a quark mass correction:
$$\delta_{mass}^{v/a} \simeq - 16 \frac{\bar{m}_q^2}{m_\tau^2},$$
and $\delta_{NP}$ describes non-perturbative 
effects in terms of incalculable expectation values
of quark and gluon operators in the operator product expansion (OPE):
$$\delta_{NP}^{v/a} \simeq
      C_4^{v/a} \frac{\vev{\cal{O}}^4}{m_\tau^4}
    + C_6^{v/a} \frac{\vev{\cal{O}}^6}{m_\tau^6}
    + C_8^{v/a} \frac{\vev{\cal{O}}^8}{m_\tau^8}.$$
The $C_n$ coefficients discribe short-distance effects,
calculable in QCD; and the expectation values for the operators
are the non-perturbative condensates. For example,
$$\vev{\cal{O}}^4 \sim \vev{\frac{\alpha_S}{\pi} GG}
              + \vev{m \bar{\psi}_q \psi_q} .$$

The important point is that one can calculate
distinct forms for $\delta_{pert}$ and $\delta_{NP}$
for each of the moments (values of $k$ and $l$),
separately for $V$ and $A$.
One can measure several different moments,
and from these, extract values for $\alpha_S(m_\tau^2)$
and for each of the non-perturbative condensates.
The result depends only on the method used to obtain
the QCD perturbation expansion; several methods are available,
including the CIPT mentioned in section \ref{s-leptonic}.

Both OPAL and ALEPH measure the moments of their
quasi-inclusive spectral functions, and fit to extract
values for $\alpha_S(m_\tau^2)$
and for the non-perturbative condensates.
The results are presented in \cite{ref:Menke,ref:Hoecker}.
The value of $\alpha_S(m_\tau^2)$ is in good agreement
with the one determined solely from the electronic branching fraction
(section \ref{s-leptonic}), but without the assumption that
$\delta_{NP}$ is small. It extrapolates to a value
at the $Z^0$ pole, $\alpha_S(m_Z^2)$, which agrees well
with measurements made there from hadronic event shapes
and other methods.
More importantly, the non-perturbative condensates
indeed are measured to be small ($\sim 10^{-2}$).

%---------------------------------------------
\subsection{QCD Chiral Sum Rules}

One can use the structure of QCD, and/or chiral perturbation theory,
to predict the moments of the difference
$v(s)-a(s)$ of the spectral functions (with $s = q^2$).
The physics of these sum rules is reviewed in \cite{ref:Rafael}.
Four sum rules have been studied with tau decay data:
\begin{itemize}
\item First Weinberg sum rule:
$$\frac{1}{4\pi^2}
  \int_0^\infty ds \left(v_1(s) - a_1(s)\right) = f_\pi^2 $$
\item Second Weinberg sum rule:
$$\frac{1}{4\pi^2}
  \int_0^\infty ds \cdot s \left(v_1(s) - a_1(s)\right) = 0$$
\item Das-Mathur-Okubo sum rule:
$$\frac{1}{4\pi^2}
  \int_0^\infty \frac{ds}{s} 
\left(v_1(s) - a_1(s)\right) = f_\pi^2 \frac{\vev{r_\pi^2}}{3} - F_A$$
\item Isospin-violating sum rule:
\begin{eqnarray*}
\frac{1}{4\pi^2}
  \int_0^\infty ds && s \ln\frac{s}{\Lambda^2}
   \left(v_1(s) - a_1(s)\right)  =  \\
    && -\frac{16\pi^2 f_\pi^2}{3\alpha} 
   \left( m_{\pi^\pm}^2 - m_{\pi^0}^2 \right). 
\end{eqnarray*}
\end{itemize}

The first, second, and fourth sum rule listed above
have definite predictions on their right-hand side,
and the data can be used to test those predictions.
However, the spectral functions measured in tau decay
extend up to $s = m_\tau^2$, not infinity.
So in practice, the tests only allow one to address the 
question, is $m_\tau^2$ close enough to infinity; 
is it ``asymptotia''?

\begin{figure}[ht]
\centerline{
\psfig{figure=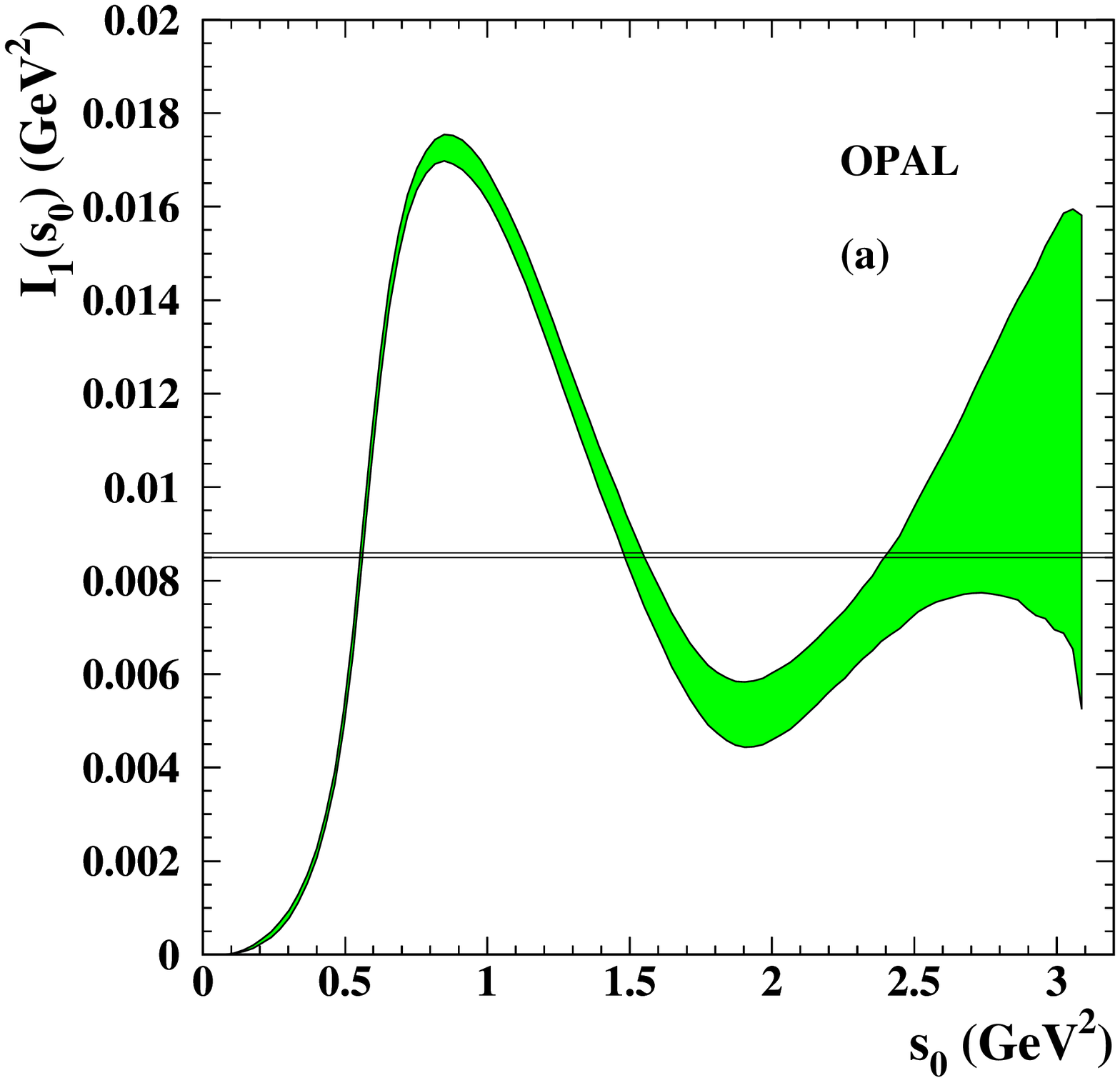,width=1.2in}
\psfig{figure=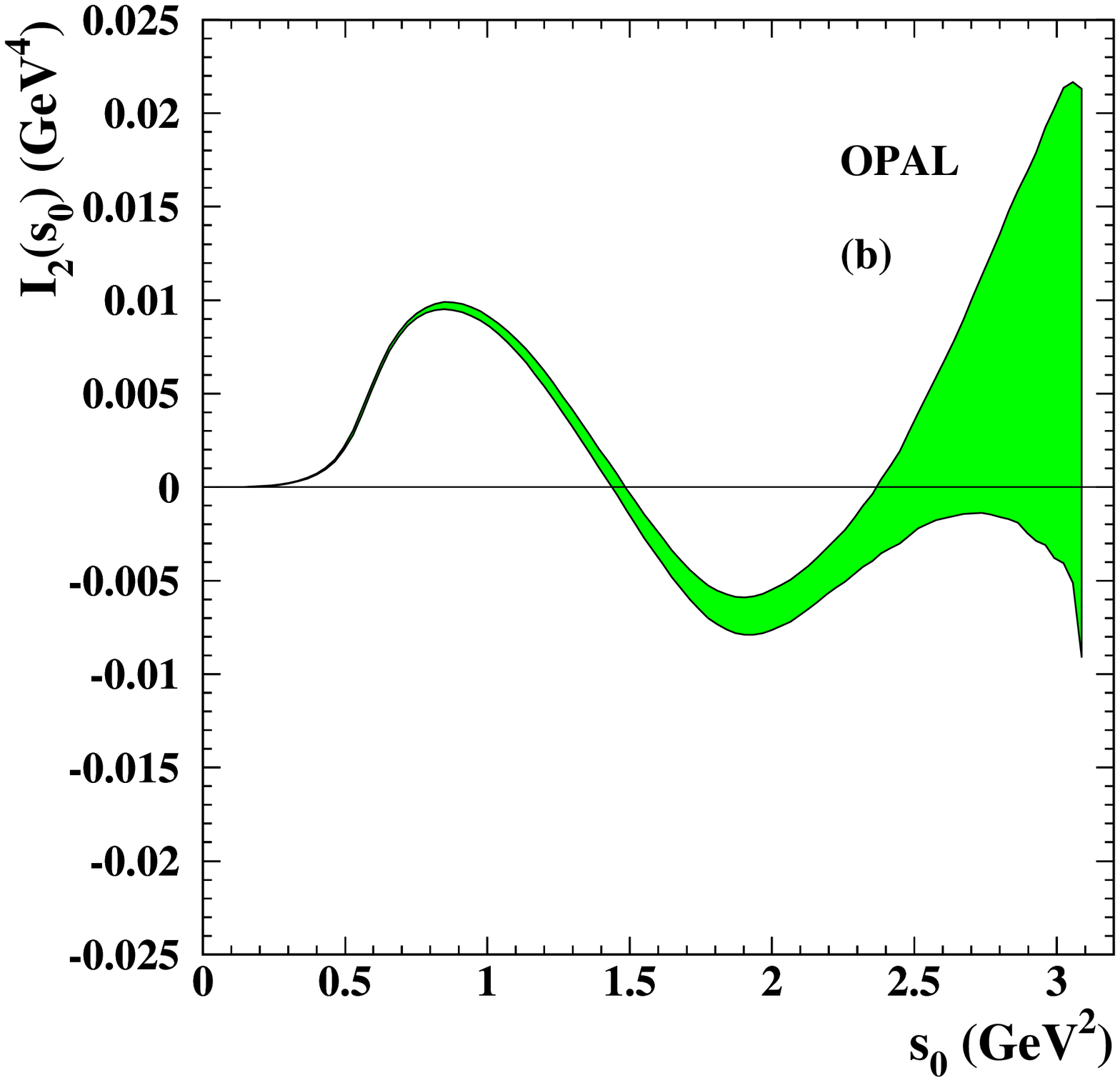,width=1.2in}}
\centerline{
\psfig{figure=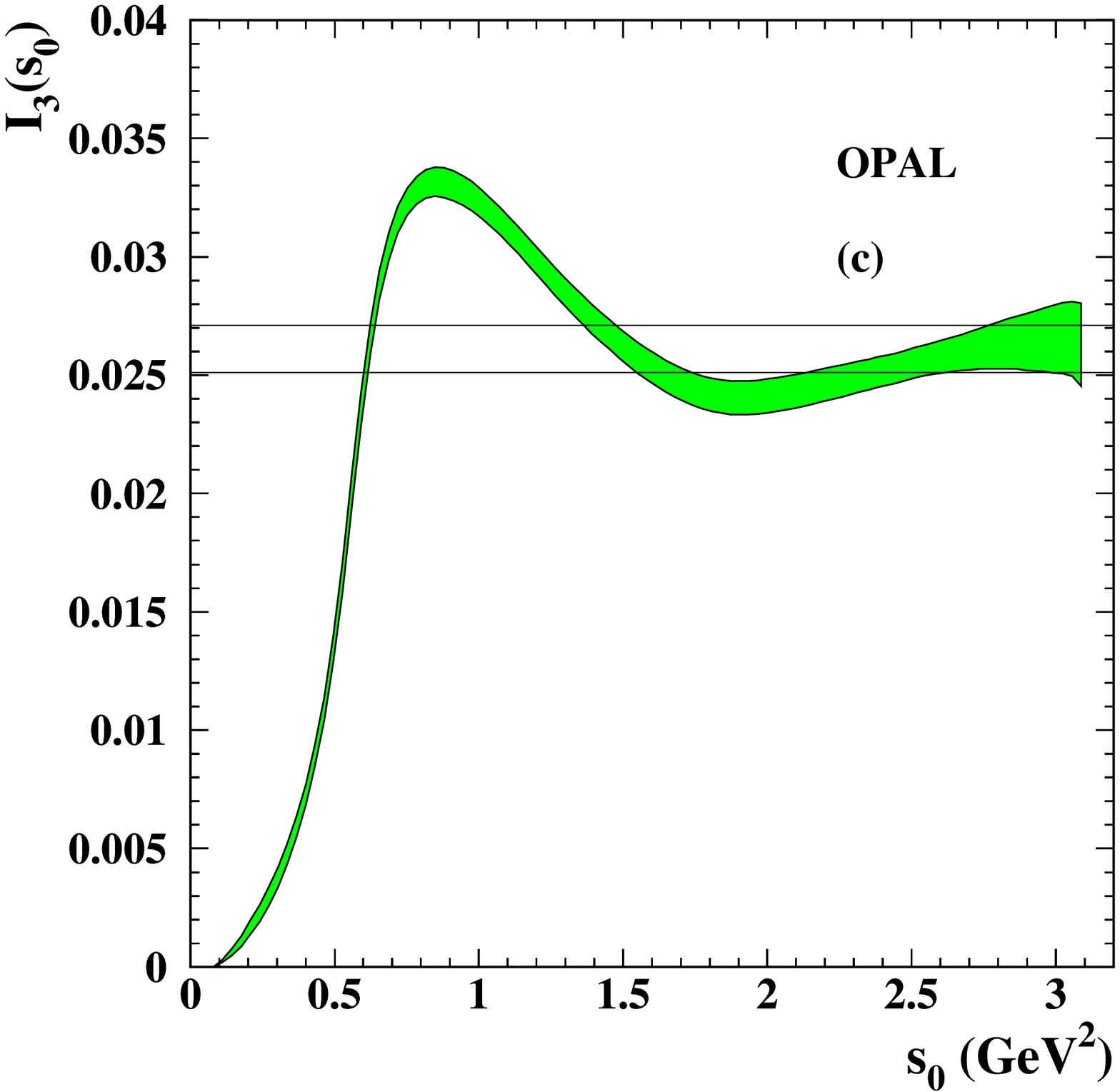,width=1.2in}
\psfig{figure=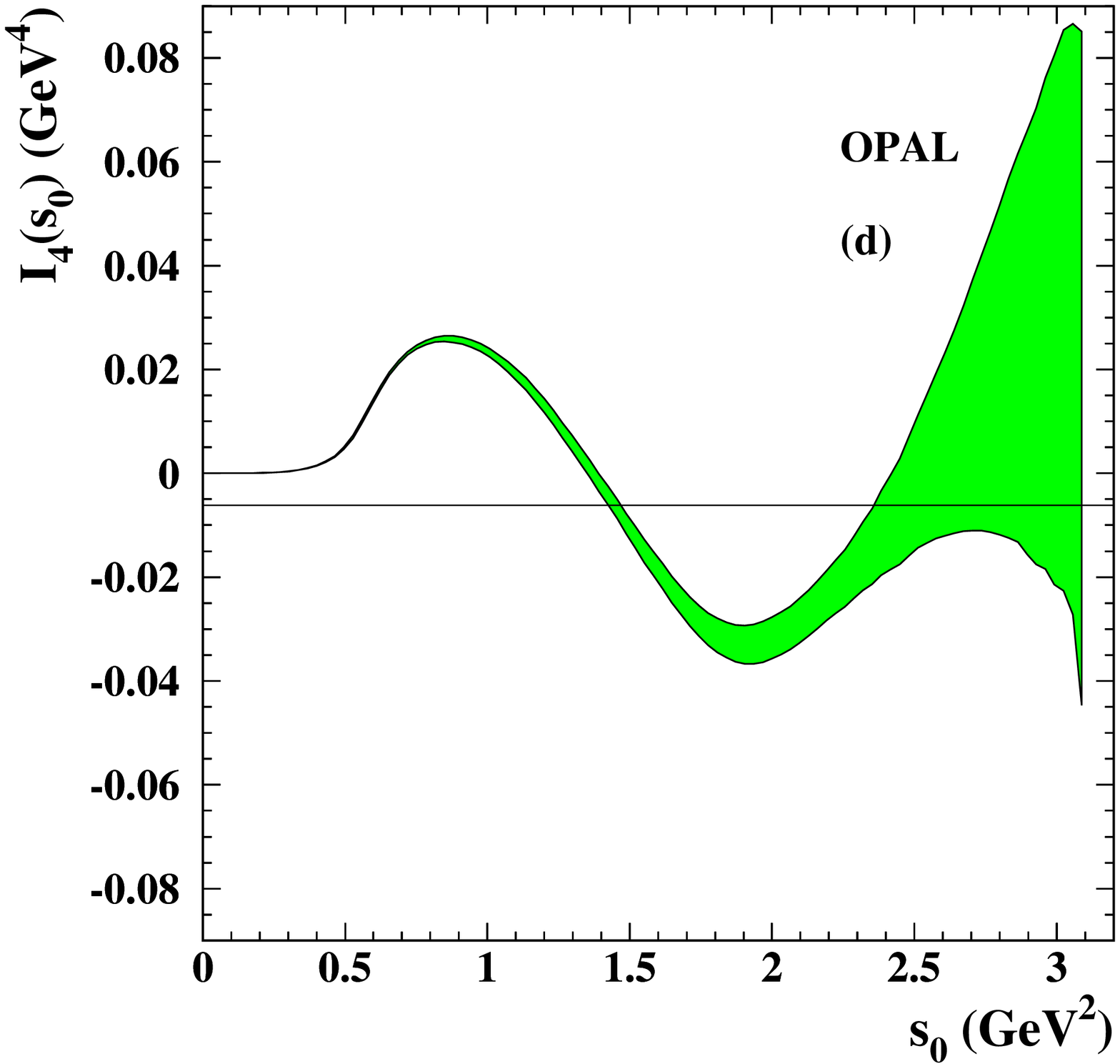,width=1.2in}}
\caption[]{QCD sum rule integrals versus the upper integration limit
from OPAL data, for the four sum rules given in the text.
The chiral prediction is given by the lines \cite{ref:Menke}.}
\label{fig:qcdsum}
\end{figure}

So far, the data are {\it consistent} with the 
sum rule predictions and with the assumption
that $m_\tau^2$ is sufficiently close to infinity
(see Fig.~\ref{fig:qcdsum} for the OPAL results);
however, the data are not yet
sufficiently precise to 
provide a quantitative test of these predictions.

However, ALEPH has studied the evolution of
the integral of the spectral functions and their moments
as a function of the cutoff $s \le m_\tau^2$,
and compared them with the theoretical prediction
for the perturbative and non-perturbative terms
as a function of their renormalization scale $s$
(fixing them at $s=m_\tau^2$ to the values they 
obtain from their fits).
In all cases, they find \cite{ref:Hoecker}
that the experimental
distributions and the theoretical predictions
overlap and track each other well 
before $s=m_\tau^2$. It appears that $s=m_\tau^2$
{\it is} asymptotia.

One can use the third (DMO) sum rule to extract
a value for the pion electric polarizability 
$$\alpha_E = \frac{\alpha F_A}{m_\pi f_\pi^2}.$$
This can be compared with predictions from
the measured value of the axial-vector form factor $F_A$,
which give $\alpha_E = (2.86\pm 0.33)\times 10^{-4}$ fm$^3$.
OPAL \cite{ref:Menke}
obtains $\alpha_E = (2.71\pm 0.88)\times 10^{-4}$ fm$^3$,
in good agreement with the prediction.

%---------------------------------------------
\subsection{$\mathbf{(g-2)_\mu}$ from $\mathbf{v(s)}$ and CVC}
\label{ss-gminus2}

As noted in section \ref{ss-emdipole},
the muon's anomalous magnetic moment $a_\mu$ is
measured with far higher precision than that of the tau,
and is in excellent agreement with the precise
theoretical prediction.
The experimental precision will soon improve considerably
\cite{ref:Grosse},
and threatens to exceed the precision with which
the theoretical prediction is determined.

Until recently, the contribution to $a_\mu$,
$$\left(\frac{g-2}{2}\right)_\mu \equiv a^\gamma_\mu
       = a_\mu^{QED} + a_\mu^{W} + a_\mu^{had} ,$$
from virtual hadronic effects $a_\mu^{had}$ had large uncertainties.
The contribution from weak effects 
($a_\mu^{W}$, from $W$-exchange vertex correction)
is small:
$a_\mu^W = (151\pm 40)\times 10^{-11}$.
Observing this contribution is one of the goals of the
current round of measurements.
(A more important goal is to look for effects of the same scale
due to physics beyond the Standard Model).
The contribution from hadronic effects
(quark loops in the photon propagator in the EM vertex correction,
and, to a lesser extent, ``light-by-light'' scattering \cite{ref:Czarnecki})
is much larger, and its uncertainty was larger than the
entire contribution from $a_\mu^W$:
$a_\mu^{had} = (7024\pm 153)\times 10^{-11}$.

This value for $a_\mu^{had}$ was obtained by relating
the quark loops in the photon propagator
to the total rate for $\gamma^*\to q\bar{q}$
as measured in $e^+e^-$ annihilation experiments at low
$s = q^2 <$ (2 GeV)$^2$.
Unfortunately, these experiments had significant 
overall errors in their measured values for the 
total hadronic cross section $\sigma(s)$.
These results are currently being improved,
as reported in \cite{ref:Eidelman}.

In the meantime, one can use the
vector part of the total decay rate 
of the tau to non-strange final states, $v(s)$,
to determine $a_\mu^{had}$.
One must assume CVC, which relates the 
vector part of the weak charged current
to the isovector part of the electromagnetic current;
and one must correct for the isoscalar part of the current
which cannot be measured in tau decay.
In addition, one must use other data
to estimate the contribution to $a_\mu^{had}$
from $s > m_\tau^2$; however, the contribution
from $s < m_\tau^2$ dominates the value and the error.

Using the vector spectral function $v(s)$
measured by ALEPH, one obtains \cite{ref:Davier}
a value for $a_\mu^{had}$ with improved errors:
$a_\mu^{had} =  (6924\pm  62)\times 10^{-11}$.
Now the error is smaller than the contribution
from $a_\mu^W$, and the sensitivity of the 
forthcoming precision experimental result
to new physics is greatly improved.

However, the use of tau data to determine 
$a_\mu^{had}$ with high precision
relies on CVC to 1\%.
Is this a valid assumption?

%---------------------------------------------
\subsection{Testing CVC}
\label{ss-cvc}

To test the validity of CVC at the per cent level,
one can compare the new
VEPP-II data on $e^+e^-\to 2n\pi$ \cite{ref:Eidelman}
to data from tau decays (from ALEPH, DELPHI, and CLEO).
When this is done, small discrepancies appear,
both in individual channels ($2\pi$ and $4\pi$)
and in the total rate via the vector current.
Discrepancies are expected, at some level, 
because of isospin violation.

These comparisons are made in \cite{ref:Eidelman},
where the data from VEPP-II are converted (using CVC)
into predictions for the branching fractions of the tau
into the analogous (isospin-rotated)
final states:
   $$ \frac{\BR(\tau\to\pi\pi\nu) - B_{CVC}}{\BR(\tau\to\pi\pi\nu)}
      = (3.2\pm 1.4)\%, $$
   $$\frac{\Delta\BR}{\BR} (2\pi+4\pi+6\pi+\eta\pi\pi+K\bar{K})
      = (3.6\pm 1.5)\%. $$
In addition, a comparison of the spectral function {\it shape}
extracted from $\tau\to 2\pi\nu_\tau$ 
and that extracted from $e^+e^-\to 2\pi$
shows discrepancies at the few per cent level.

It is not clear whether these comparisons mean that
CVC is only good to $\sim 3\%$,
or whether the precision in the data used for the comparison
needs improvement.
To be conservative, however, results that rely on
CVC should be quoted with an error
that reflects these discrepancies.

%---------------------------------------------
\section{EXCLUSIVE FINAL STATES}
\label{s-structure}

The semi-hadronic decays of the tau to
exclusive final states is the realm of low energy meson dynamics,
and as such cannot be described with perturbative QCD.
In the limit of small energy transfers, 
chiral perturbation theory can be used to predict rates;
but models and symmetry considerations must be used to 
extrapolate to the full phase space of the decay.

In general, the Lorentz structure of the decay
(in terms of the 4-vectors of the final state pions, kaons, 
and $\eta$ mesons) can be specified; models are then required
to parameterize the a priori unknown form factors
in the problem.
Information from independent measurements in low energy
meson dynamics can be used to reduce the number of free
parameters in such models; an example is given in \cite{ref:bingan}.

Measurements of the total branching fractions
to different exclusive final states
(\eg, $h n\pi^0\nutau$, $n = 0 ... 4$,
  $3h n\pi^0\nutau$, $n = 0 ... 3$,
  $5h n\pi^0\nutau$, $n = 0 ... 1$, with $h = \pi^\pm$ or $K^\pm$;
or $\eta n\pi^0\nutau$, $n = 2 ... 3$)
have been refined for many years,
and remain important work;
recent results from DELPHI are presented in \cite{ref:Lopez}.
Branching fractions for final states containing kaons
($K^\pm$, $K^0_S$, and $K^0_L$)
are presented in \cite{ref:Kravchenko,ref:Andreazza,ref:Chen}
and are discussed in more detail in section \ref{s-kaons}.
The world average summaries of all the 
semi-hadronic exclusive branching fractions
are reviewed in \cite{ref:Heltsley}.

Since the world average branching fractions for all 
exclusive tau decays now sum to one with small errors,
emphasis has shifted to the detailed study of
the structure of exclusive final states.
At previous tau workshops, the focus was on the simplest
final states with structure: $\pi\pi^0\nutau$ and $K\pi\nutau$
\cite{ref:rhostructure}.
At this workshop, the attention has shifted to the
$3\pi$, $K\pi\pi$, and $K\bar{K}\pi$ final states,
which proceed dominantly through the axial-vector current.

The results for the $K\pi\pi$, and $K\bar{K}\pi$ final states
are discussed in section \ref{s-kaons};
here we focus on $3\pi$.
The $4\pi$ final state remains to be studied in detail.
Final states with 5 or 6 pions
contain so much resonant substructure,
and are so rare in tau decays, that detailed
fits to models have not yet been attempted.
However, isospin can be used to characterize 
the pattern of decays; this is discussed,
using data from CLEO, in \cite{ref:Gan}.

Recent results on $\tau\to 3\pi\nutau$ from
OPAL, DELPHI, and CLEO are discussed in \cite{ref:Schmidtler}.
There are two complementary approaches that can be taken:
describing the decay in a Lorentz-invariant way,
parameterized by form-factors which model 
intermediate resonances; or via model-independent
structure functions, defined in a specified angular basis.
The model-dependent approach gives a simple picture 
in terms of well-defined decay chains, such as
$a_1 \to \rho\pi\to 3\pi$ or $K_1\to (K^*\pi, K\rho) \to K\pi\pi$;
but the description is only as good as the model,
and any model is bound to be incomplete.
The structure function approach results in large tables of numbers
which are harder to interpret (without a model);
but it has the advantage that some of the functions,
if non-zero, provide model-independent evidence for sub-dominant processes
such as pseudoscalar currents (\eg, $\pi^\prime(1300)\to 3\pi$)
or vector currents (\eg, $K^{*\prime} \to K\pi\pi$).

OPAL and DELPHI present fits of their $3\pi$ to two simple models
for $a_1\to \rho\pi$,
neither of which describe the data in detail \cite{ref:Schmidtler}.
DELPHI finds that in order to fit the high $m_{3\pi}$ region
with either model, a radially-excited $a_1^\prime(1700)$ meson
is required, with a branching fraction
${\cal B}(\tau\to a_1^\prime\nutau\to 3\pi\nutau)$
of a few $\times 10^{-3}$, depending upon model.
Even then, the fits to the Dalitz plot variables are poor.
The presence of an enhancement at high mass
(over the simple models) has important consequences
for the extraction of the tau neutrino mass using
$3\pi\nutau$ events.

OPAL also analyzes their data in terms of structure functions,
and from these, they set limits on scalar currents:
$$\Gamma^{scalar}/\Gamma^{tot} (3\pi\nutau) < 0.84\%,$$
and make a model-independent determination of the {\it signed}
tau neutrino helicity:
$$h_{\nu_\tau} = -1.29\pm 0.26\pm 0.11$$
(in the Standard Model, $h_{\nu_\tau} = -1$).

CLEO does a model-dependent fit to their 
$\tau^-\to\pi^-\pi^0\pi^0\nutau$ data.
They have roughly 5 times the statistics 
of OPAL or DELPHI.
This allows them to consider contributions
from many sub-dominant processes, including:
$a_1\to \rho^\prime\pi$, both S-wave and D-wave;
$a_1\to f_2(1275)\pi$, $\sigma\pi$, and $f_0(1370)\pi$;
and $\pi^\prime(1300)\to 3\pi$.
Here, the $\sigma$ is a broad scalar resonance
which is intended to ``mock up'' the complex structure
in the S-wave $\pi\pi$ scattering amplitude above threshold,
according to the Unitarized Quark Model.
CLEO also considers the process $a_1\to K^*K$,
as a contribution to the total width and therefore
the Breit Wigner propagator for the $a_1$
(of course, the final state that is studied
does not receive contributions from $K^*K$).

CLEO finds significant contributions from all of these processes,
with the exception of $\pi^\prime(1300)\to 3\pi$.
All measures of goodness-of-fit are excellent, 
throughout the phase space for the decay.
There is also excellent agreement with the
data in the $\pi^-\pi^+\pi^-\nutau$ final state,
which, because of the presence of isoscalars
in the substructure, is non-trivial.
There is strong evidence for a $K^*K$ threshold.
There is only very weak evidence for an $a_1^\prime(1700)$.
They measure the radius of the $a_1$ meson to be $\approx 0.7$ fm.
They set the 90\%\ C.L. limit 
$$\Gamma(\pi^\prime(1300)\to\rho\pi)/\Gamma(3\pi) < 1.0\times 10^{-4},$$
and make a model-dependent determination of the signed
tau neutrino helicity:
$$h_{\nu_\tau} = -1.02\pm 0.13\pm 0.03 \q(model).$$
{\it All} of these results are model-dependent;
but the model fits the data quite well.

%---------------------------------------------
\section{KAONS IN TAU DECAY}
\label{s-kaons} 

Kaons are relatively rare in tau decay,
and modes beyond $K\nutau$ and $K^*\nutau$ 
are only being measured with some precision
in recent years.
At TAU 98, ALEPH presented \cite{ref:Chen}
branching fractions for 27 distinct modes with $K^\pm$, $K^0_S$, 
and/or $K^0_L$ mesons, including $K3\pi$;
DELPHI presented \cite{ref:Andreazza} 
12 new (preliminary) branching fractions,
and CLEO presented \cite{ref:Kravchenko}
an analysis of
four modes of the form $K^- h^+\pi^-(\pi^0)\nu_\tau$.

In the $K\pi$ system, ALEPH
sees a hint of $K^{*\prime}(1410)$,
with an amplitude (relative to $K^*(892)$)
which is in good agreement with 
the analogous quantity from $\tau\to (\rho,\rho^\prime)\nutau$.
CLEO sees no evidence for anything beyond the $K^*(892)$.

ALEPH and CLEO both study the $K\pi\pi$ system.
Here, one expects contributions from:
the axial-vector $K_1(1270)$, which decays to $K^*\pi$, $K\rho$,
and other final states;
the axial-vector $K_1(1400)$, which decays predominately to $K^*\pi$;
and, to a much lesser extent, the vector $K^{*\prime}$,
via the Wess-Zumino parity-flip mechanism.
Both ALEPH and CLEO see more $K_1 (1270)$ than $K_1(1400)$,
with significant signals for $K\rho$ as well as $K^*\pi$
in the Dalitz plot projections.

The two $K_1$ resonances are quantum mechanical mixtures
of the $K_{1a}$ (from the $J^{PC} = 1^{++}$ nonet, the strange 
analog of the $a_1$),
and the $K_{1b}$ (from the $J^{PC} = 1^{+-}$ nonet, the strange 
analog of the $b_1$).
The coupling of the $b_1$ to the $W$ is a second-class current,
permitted in the Standard Model only via isospin violation.
The coupling of the $K_{1b}$ to the $W$ is permitted
only via $SU(3)_f$ violation.

CLEO extracts the $K_{1a} - K_{1b}$ mixing angle 
(with a two-fold ambiguity)
and $SU(3)_f$-violation parameter $\delta$ in $\tau \to K_{1b}\nu_\tau$,
giving results consistent with 
previous determinations from hadroproduction experiments
\cite{ref:Kravchenko}.

ALEPH studies the $K\bar{K}\pi$ structure, and finds \cite{ref:Chen}
that $K^*K$ is dominant, with little contribution
from $\rho\pi$, $\rho\to K\bar{K}$.
The $K\bar{K}\pi$ mass spectrum is consistent 
with coming entirely from $a_1\to K\bar{K}\pi$,
although there may be a large vector component.

\begin{figure}[ht]
\psfig{figure=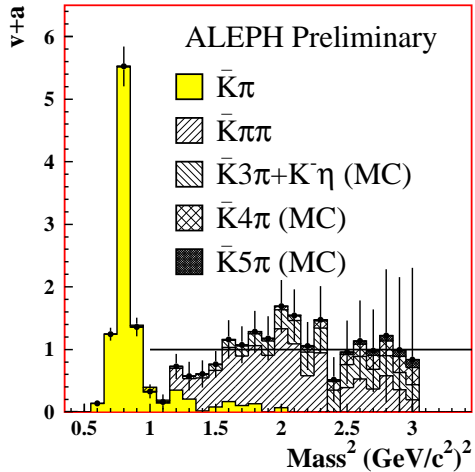,width=2.6in}
\caption[]{Total $V+A$ spectral function from $\tau$ decays into
 strange final states. from ALEPH.
The contributions from the exclusive channels, from data and MC,
are indicated \cite{ref:Chen}.}
\label{fig:specs_aleph}
\end{figure}

ALEPH also analyzes the isospin content of the
$K\pi$, $K\pi\pi$, and $K\bar{K}\pi$
systems. Finally, they classify the net-strange final states
as arising from the vector or axialvector current,
and construct the strange spectral function
$(v+a)^S_1(s)$ (using the data for the $K\pi$ and $K\pi\pi$
components, and the Monte Carlo for the small contributions
from $K3\pi$, $K4\pi$, \etc)  \cite{ref:Chen}.
This function, shown in Fig.~\ref{fig:specs_aleph},
can then be used for QCD studies, as discussed
in the next section.

%---------------------------------------------
\subsection{$\mathbf{m_s}$ from $\mathbf{R_{\tau,s}}$}
\label{ss-ms}

The total strange spectral function can be used
to extract QCD parameters, in direct analogy
with the total and non-strange rates and moments
as described in sections \ref{s-leptonic} and \ref{ss-moments}.
In the strange case, we have
\begin{eqnarray*}
R_\tau^s &\equiv& \frac{\BR_{Kn\pi}}{\BR_e} \\
       &=& 3 V_{us}^2 S_{EW} (1+\delta_{pert}+\delta_{mass}^s+\delta_{NP}),
\end{eqnarray*}
and we focus on the quark mass term:
$$\delta_{mass}^s \simeq 
   - 8 \frac{\bar{m}_s^2}{m_\tau^2} \left[ 
    1 +\frac{16}{3}\frac{\alpha_S}{\pi} + 
   {\cal O}\left(\frac{\alpha_S}{\pi}\right)^2 \right],$$
where $\bar{m}_s = m_s(m_\tau^2)$ is the $\overline{\mbox{MS}}$
running strange quark mass, evaluated at the tau mass scale.
For $m_s(m_\tau^2) \approx 150$ MeV/c$^2$, we expect
$\delta_{mass}^s \approx -10\%$, and $R_\tau^s \approx 0.16$
(but with a large uncertainty due to poor convergence of 
the QCD expansion).

The history of the calculation of the 
${\cal O}\left(\frac{\alpha_S}{\pi}\right)^2$ term,
and the apparent convergence of the series, has been rocky.
But considerable progress has been made in the last year,
and the theoretical progress is reviewed in \cite{ref:Prades}.

The value of the strange quark mass appears in many 
predictions of kaon properties;
most importantly, it appears in the theoretical expression
for the parameter governing direct CP violation
in the kaon system, $\epsilon^\prime/\epsilon$.
Thus we need to know its value in order to extract information
on the CKM matrix from measurements of direct CP violation.

ALEPH has constructed the strange spectral function
as described in the previous section
and shown in Fig.~\ref{fig:specs_aleph},
and has calculated its integral:
$$R_\tau^s = 0.1607\pm0.0066$$
and its moments $R_{\tau,s}^{kl}$.
By comparing with the non-strange moments,
they cancel, to lowest order,
the mass-independent non-perturbative terms.
They fit for $m_s(m_\tau^2)$
and the residual non-perturbative condensates.
They obtain \cite{ref:Hoecker}
the strange quark running mass
$m_s(m_\tau^2) = (163^{+34}_{-43})$ MeV/c$^2$.
At the (1 GeV)$^2$ scale, 
$m_s(1\,\mbox{GeV}^2) = (217^{+45}_{-57})$ MeV/c$^2$,
which compares well with estimates from 
sum rules and lattice calculations.
The uncertainty is rather large, especially
the component that comes from the uncertainty
in the convergence in the QCD series;
but improvements are expected.

%---------------------------------------------
\section{TAU NEUTRINO MASS}   %{$m_{\nu_\tau}$}
\label{s-mnutau}

In the Standard Model, the neutrinos are assumed to be massless,
but nothing prevents them from having a mass. 
If they do, they can mix (in analogy with the down-type quarks),
exhibit CP violation, and potentially decay.
In the so-called ``see-saw'' mechanism, the tau neutrino
is expected to be the most massive neutrino.

Indirect bounds from cosmology and big-bang nucleosynthesis
imply that
if the $\nu_\tau$ has a lifetime long enough to have not decayed
before the period of decoupling,
then a mass region between 65 eV/c$^2$ and 4.2 GeV/$c^2$ 
can be excluded.
For unstable neutrinos, these bounds can be evaded.
A massive Dirac neutrino will have a right-handed component,
which will interact very weakly with matter via the 
standard charged weak current.
Produced in supernova explosions, these right-handed neutrinos
will efficiently cool the newly-forming neutron star,
distorting the time-dependent neutrino flux.
Analyses of the detected neutrino flux
from supernova 1987a,
results in allowed ranges $m_{\nu_\tau} < 15-30$ KeV/$c^2$
or $\mnutau > 10-30$ MeV/$c^2$,
depending on assumptions.
This leaves open a window for an MeV-range mass for $\nu_\tau$
of 10-30 MeV/$c^2$, with lifetimes on the order of 
$10^5 - 10^9$ seconds.

The results from Super-K 
(\cite{ref:Nakahata}, see section \ref{s-neuosc} below)
suggest neutrino mixing, and therefore, mass.
If they are observing $\nu_\mu \lra \nu_\tau$
oscillation, then \cite{ref:McNulty} $m_{\nu_\tau} < 170$ KeV/c$^2$,
too low to be seen in collider experiments.
If instead they are observing oscillations
of $\nu_\mu$ to some sterile neutrino,
then there is no information from Super-K
on the tau neutrino mass.

If neutrino signals are observed from a galactic supernova,
it is estimated that neutrino masses as low as of 25 eV/$c^2$
could be probed by studying the dispersion in arrival time
of the neutrino events in a large underground detector
capable of recording neutral current interactions.
Very energetic neutrinos from a distant active galactic nucleus (AGN)
could be detected at large underground detectors (existing or planned).
If neutrinos have mass and therefore a magnetic moment,
their spin can flip in the strong magnetic field of the AGN,
leading to enhanced spin-flip and flavor oscillation effects 
\cite{ref:Husain}.
The detectors must have the ability to measure the
direction of the source, the energy of the neutrino,
and its flavor.

\begin{figure}[ht]
\psfig{figure=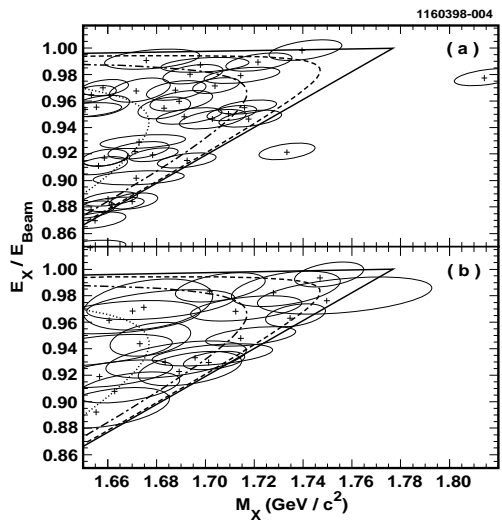,width=2.6in}
\caption[]{The scaled hadronic energy $E_{n\pi}^{lab}/E_{beam}$
\vs\ hadronic mass for (a) the $5\pi$ and
(b) the $3\pi^\pm2\pi^0$ event candidates from CLEO.
Ellipses represent the resolution countours \cite{ref:Duboscq}.}
\label{fig:duboscq1}
\end{figure}

At TAU 98, CLEO presented \cite{ref:Duboscq} two new limits,
one using $\tau\to 5\pi^\pm\nutau$ and $3\pi^\pm2\pi^0\nutau$,
and a preliminary result using $3\pi^\pm\pi^0\nutau$.
Both results are based on the distribution of events in the
2-dimensional space of $m_{n\pi}$ \vs\ $E_{n\pi}^{lab}$,
looking for a kinematical suppression in the 
high-mass, high energy corner
due to a 10 MeV-scale neutrino mass (see Fig.~\ref{fig:duboscq1}).
This technique has been used previously by ALEPH, DELPHI, and OPAL.

A summary of the best direct tau neutrino mass limits 
at 95\%\ C.L.~is given \cite{ref:McNulty} in Table~\ref{tab:numass}.

\begin{table}[!ht]
\centering
\label{tab:numass}
\caption[]{Limits on the $\tau$ neutrino mass.}
  \begin{tabular}{llr} \hline
   Indirect & from ${\cal B}_e$ & 38 MeV \\
   ALEPH & $5\pi(\pi^0)$   & 23 MeV \\
   ALEPH & $3\pi$          & 30 MeV \\
   ALEPH & both            & 18.2 MeV \\
   OPAL  & $5\pi$          & 43 MeV \\
   DELPHI & $3\pi$         & 28 MeV \\
   OPAL   & $3\pi - vs - 3\pi$ & 35 MeV \\
   CLEO (98)  & $5\pi$, $3\pi2\pi^0$ & 30 MeV \\
   CLEO (98p) & $4\pi$               & 31 MeV \\
  \hline
  \end{tabular}
\end{table}

The limits are usually dominated by a few ``lucky'' events
near the endpoint, which necessarily have a low probability;
they are likely to be upward fluctuations
in the detector's mass \vs\ energy resolution response.
Therefore, it is essential to understand and model 
that response, especially the tails.
Extracting meaningful limits on the neutrino mass
using such kinematical methods is made difficult by
many subtle issues regarding resolution,
event migration, modeling of the spectral functions,
and certainly also {\it luck}.
Pushing the discovery potential to or below 10 MeV
will require careful attention to these issues,
but most importantly, much higher statistics.
We look forward to the high-statistics event samples
obtainable at the B Factories soon to come on line.

%---------------------------------------------
\section{NEUTRINO OSCILLATIONS}
\label{s-neuosc}

If neutrinos have mass, 
then they can mix with one another,
thereby violating lepton family number conservation.
In a two-flavor oscillation situation,
a beam of neutrinos that are initially of one pure flavor, 
\eg, $\nu_\mu$
will oscillate to another flavor, \eg, $\nu_\tau$
with a probability given by
$$P(\nu_\mu\to\nu_\tau) = 
\sin^2(2\theta_{\mu\tau})\sin^2(\pi L/L_0) ,$$
where the strength of the oscillation is governed by
the mixing parameter $\sin^2 2\theta_{\mu\tau}$.
$L$ is the distance from the source of initially pure $\nu_\mu$
in meters, and
the oscillation length $L_0$ is given by
$$ L_0 = \frac{2.48 E_\nu\,[\mbox{GeV}]}{
              \Delta m^2\,[\mbox{eV}^2] }. $$
$\Delta m^2$ is the difference of squared masses 
of the two flavors measured in eV$^2$, and
$E_\nu$ is the neutrino energy measured in GeV.
This formula is only correct in vacuum.
The oscillations are enhanced if the neutrinos are travelling
through dense matter, as is the case for neutrinos 
from the core of the Sun. 
This enhancement, known as the MSW effect, is invoked
as an explanation
for the deficit of $\nu_e$ from the Sun's
core as observed on Earth.
In such a scenario, all three neutrino flavors mix with one another.

Evidence for neutrino oscillations
has been seen in the solar neutrino deficit
($\nu_e$ disappearance),
atmospheric neutrinos (apparently, $\nu_\mu$ disappearance),
and neutrinos from $\mu$ decay ($\nu_\mu\to \nu_e$ 
and $\bar{\nu}_\mu\to \bar{\nu}_e$ appearance, at LSND).
At this workshop, upper limits were presented
for neutrino oscillations from the $\nu_\mu$ beam at CERN, from
NOMAD \cite{ref:Paul} and CHORUS \cite{ref:Cussans}.

It appears that, if the evidence for neutrino oscillations
from solar, atmospheric, and LSND experiments are {\it all} correct,
the pattern cannot be explained with only three
Standard Model Dirac neutrinos.
There are apparently three distinct $\Delta m^2$ regions:
\begin{eqnarray*}
\Delta m^2_{solar} &=& 10^{-5} \ \mbox{or}\ 10^{-10}\ \mbox{eV}^2 \\
\Delta m^2_{atmos} &=& 10^{-2} \ \mbox{to}\ 10^{-4}   \mbox{eV}^2 \\
\Delta m^2_{LSND}  &=& 0.2     \ \mbox{to}\ 2       \ \mbox{eV}^2 .
\end{eqnarray*}
This mass hierarchy is difficult (but not impossible)
to accomodate in a 3 generation model.
The addition of a 
$4^{th}$ (sterile? very massive?) neutrino 
can be used to describe either the solar neutrino
or atmospheric neutrino data \cite{ref:Gonzalez}
(the LSND result requires $\nu_\mu\to \nu_e$).
The introduction of such a $4^{th}$ neutrino makes it 
relatively easy to describe all the data.
In addition, a light sterile neutrino is a candidate
for hot dark matter, and a heavy neutrino, for cold dark matter.

\subsection{Results from Super-K}
\label{ss-superk}

We turn now to the results on neutrino oscillations
from Super-Kamiokande, certainly 
the highlight of {\it any} physics conference in 1998. 

\begin{figure}[ht]
\psfig{figure=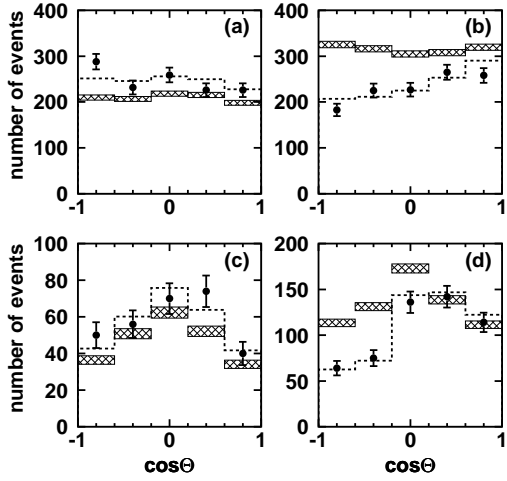,width=2.6in}
\caption[]{Zenith angle distribution of atmospheric
neutrino events: 
(a) sub-GeV $e$-like;
(b) sub-GeV $\mu$-like;
(c) multi-GeV $e$-like;
(d) multi-GeV $\mu$-like and partially-contained.
Shaded histograms give the MC expectations
without oscillations, dotted histograms
show the best fit assuming neutrino oscillations \cite{ref:Nakahata}.}
\label{fig:superkz}
\end{figure}

Neutrinos produced in atmospheric cosmic ray showers
should arrive at or below the surface of the earth
in the ratio $(\nu_\mu+\bar{\nu}_\mu)/(\nu_e+\bar{\nu}_e) \simeq 2$
for neutrino energies $E_\nu < 1$ GeV, and somewhat higher
at higher energies. There is some uncertainty in the 
flux of neutrinos of each flavor from the atmosphere,
so Super-K measures \cite{ref:Nakahata}
the double ratio:
$$ R = 
  \left(\frac{\nu_\mu+\bar{\nu}_\mu}{\nu_e+\bar{\nu}_e}\right)_{observed}
  \left/
  \left(\frac{\nu_\mu+\bar{\nu}_\mu}{\nu_e+\bar{\nu}_e}\right)_{calculated}.
  \right.
$$
They also measure the 
zenith-angle dependence of the flavor ratio;
upward-going 
neutrinos have traveled much longer since they were produced,
and therefore have more time to oscillate.
Super-K analyzes several classes of events
(fully contained and partially contained,
sub-GeV and multi-GeV),
classifying them as ``e-like'' and ``$\mu$-like''.
Based on the flavor ratio, the double-ratio, 
the lepton energy, 
and the zenith-angle dependence, 
they conclude that they are observing effects
consistent with $\nu_\mu$ disappearance and thus neutrino oscillations
(see Fig.~\ref{fig:superkz}).
Assuming $\nu_\mu\to\nu_\tau$, their best fit gives
$\Delta m^2 = 2.2\times 10^{-3}\,\mbox{eV}^2$, $\sin^2 2\theta_{\mu\tau} = 1$
(see Fig.~\ref{fig:icarustau}).
They also measure the 
upward through-going and stopping muon event rates
as a function of zenith angle, and see consistent results.
Whether these observations imply that muon neutrinos are oscillating
into tau neutrinos or into some other (presumably sterile) neutrino
is uncertain.

Super-K also observes \cite{ref:Nakahata}
some 7000 low-energy $\nu_e$ events
which point back to the sun, presumably produced during
$^8B$ synthesis in the sun.
They measure a flux which is significantly smaller 
than the predictions from standard solar models
with no neutrino mixing
($\approx 40\%$, depending on the model),
with a hint of energy dependence in the (data/model) ratio.
They see no significant difference
between solar neutrinos which pass through the earth
(detected at night) and those detected during the day.
% Their allowed region is shown in Fig.~\ref{fig:superk}.

\subsection{NOMAD and CHORUS}
\label{ss-nomad}

Two new accelerator-based 
``short-baseline'' neutrino oscillation experiments
are reporting null results at this workshop.
The NOMAD and CHORUS detectors are situated in the
$\nu_\mu$ beam at CERN, searching for $\nu_\mu\to \nu_\tau$ oscillations.
They have an irreducible background from $\nu_\tau$ in the beam
(from $D_s$ production and decay),
but it is suppressed relative to the $\nu_\mu$ flux by $5\times 10^{-6}$.

\begin{figure}[ht]
\psfig{figure=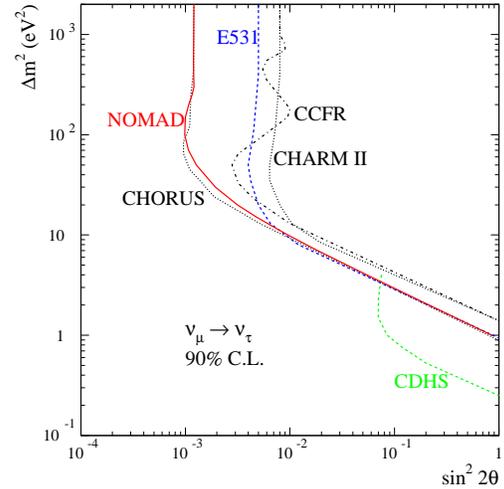,width=2.6in}
\caption[]{Exclusion plot at 90\%\ C.L.~for $\nu_\mu\lra\nu_\tau$
oscillations, from CHORUS and NOMAD \cite{ref:Paul}.}
\label{fig:nomad}
\end{figure}

NOMAD is an electronic detector which searches for
tau decays to $e\nu\nu$, $\mu\nu\nu$,
$h n\pi^0\nu$, and $3\pi n\pi^0\nu$.
They identify them as decay products of a tau
from kinematical measurements,
primarily the requirement of
missing $p_t$ due to the neutrino(s).
They see no evidence for oscillations \cite{ref:Paul},
and set the following limits at 90\%\ C.L.:
   $$P(\nu_\mu \to \nu_\tau) < 0.6\times 10^{-3}$$
   $$\sin^22\theta_{\mu\tau} < 1.2\times 10^{-3}$$
     for $\Delta m^2 > 10^2$ eV$^2$.
The exclusion plot is shown in Fig.~\ref{fig:nomad}.

The CHORUS experiment triggers on
oscillation candidates with an electronic detector,
then aims for the direct observation of production and decay
of a tau in a massive nuclear emulsion stack target.
they look for tau decays to $\mu\nu\nu$ or $h n\pi^0\nu$,
then look for a characteristic pattern in their emulsion
corresponding to: nothing, then a 
tau track, a kink, and a decay particle track.
They also see no evidence for oscillations \cite{ref:Cussans},
and set limits that are virtually identical to those of NOMAD;
the exclusion plot is shown in Fig.~\ref{fig:nomad}.

\begin{figure}[ht]
\psfig{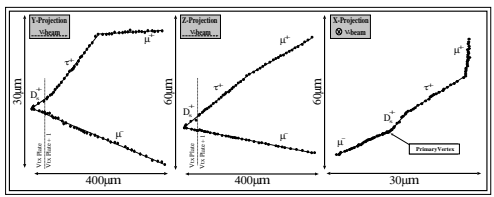}
\caption[]{Tracks in the CHORUS emulsion 
from a $D_s\to \tau\nu$ candidate \cite{ref:Cussans}.}
\label{fig:chorusevt}
\end{figure}

CHORUS has also reconstructed a beautiful event,
shown in Fig.~\ref{fig:chorusevt},
in which a $\tau^-$ lepton is tracked 
in their emulsion \cite{ref:Cussans}.
They infer the decay chain:
$\nu_\mu N \to \mu^- D^{*+}_s N$,
$D^{*+}_s \to D^{+}_s \gamma$,
$D^{+}_s \to \tau^+\nu_\tau$,
$\tau^+\to \mu^+\nu_\mu\bar{\nu}_\tau$.
In their emulsion, they track the $D^{+}_s$, the $\mu^-$,
the $\tau^+$, and the decay $\mu^+$.

\subsection{OBSERVATION OF $\mathbf{\nu_\tau}$}

The appearance of a $\tau$ and its subsequent decay
in a detector exposed to a neutrino beam
would constitute direct observation
of the tau neutrino.
Fermilab experiment 872 ({\bf D}irect {\bf O}bservation
of {\bf NU\_T}au, DONUT) is designed to directly see
for the first time, such events.
The experiment relies on the production of $D^+_s$
mesons, which decay to $\tau^+\nu_\tau$ with 
branching fraction $\simeq 4\%$.
A detector downstream from a beam dump
searches for $\tau^-$ production and subsequent
decay (with a kink) in an emulsion target.
They have collected, and are presently analyzing, their data;
they expect to find $40\pm12$ $\nu_\tau$ interactions \cite{ref:Thomas}.

\begin{figure}[ht]
\psfig{figure=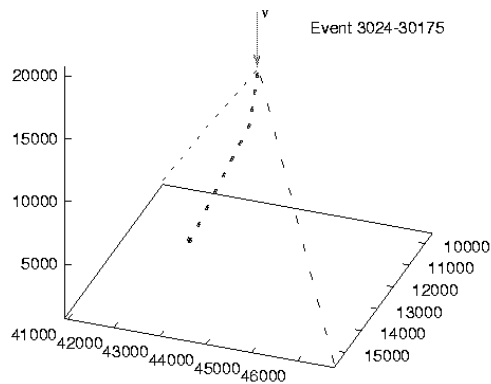,width=2.6in}
\caption[]{Candidate event for $\nu_\tau\to \tau X$, $\tau\to \mu\nu\nu$,
in the DONUT emulsion. 
There is a 100 mrad kink 4.5 mm from the interaction vertex.
The scale units are microns \cite{ref:Thomas}.}
\label{fig:donutevt}
\end{figure}

At TAU 98, they showed \cite{ref:Thomas} an event 
consistent with such an interaction; see Fig.~\ref{fig:donutevt}.
If/when a sufficient number of such events are
observed and studied, there will finally be
direct observational evidence for the tau neutrino.

%---------------------------------------------
\section{WHAT NEXT?}
\label{s-whatnext}

The frontiers of tau physics center on ever-higher precision,
farther reach for rare and forbidden phenomena,
and a deeper study of neutrino physics.
The next generation of accelerator-based
experiments will search for
neutrino oscillations with the small $\Delta m^2$
suggested by the Super-K results,
and do precision studies of $\tau$ decays
with samples exceeding $10^8$ events.

\subsection{Long-baseline neutrino oscillations}
\label{ss-longbase}

Several ``long-baseline'' experiments are planned,
in which $\nu_\mu$ beams are produced at accelerators,
allowed to drift (and hopefully, oscillate
into $\nu_e$, $\nu_\tau$, or sterile neutrinos)
for many km, and are then detected.
The motivation comes from the Super-K results,
which suggest $\nu_\mu \lra \nu_\tau$ oscillations
with $\Delta m^2$ in the $10^{-2} - 10^{-3}$ eV$^2$ range.
For such small $\Delta m^2$ values,
accelerator-produced $\nu_\mu$ beams must travel many kilometers
in order to get appreciable $\nu_\mu\to \nu_\tau$ conversion.

At Fermilab, an intense, broad-band $\nu_\mu$ beam
is under construction (NuMI).
The MINOS experiment consists of 
two nearly identical detectors which will search for
neutrino oscillations with this beam \cite{ref:Thomas}.
The near detector will be on the Fermilab site.
The neutrino beam will pass through Wisconsin, 
and neutrino events will be detected
at a far detector 
in the Soudan mine in Minnesota, 720 km away.
The experiment is approved, and is scheduled
to turn on in 2002.

From the disappearance of $\nu_\mu$'s between the
near and far detectors, 
MINOS will be sensitive \cite{ref:Thomas}
to oscillations with mixing angle
$\sin^2 2\theta \gsim 0.1$
for $\Delta m^2 \gsim 2\times 10^{-3}$ eV$^2$.
From the appearance of an excess of $\nu_e$ events
in the far detector, they are sensitive to 
$\nu_\mu\to\nu_e$ mixing down to 
$\sin^2 2\theta \gsim 0.002$
for $\Delta m^2 \gsim 2\times 10^{-2}$ eV$^2$.
MINOS may be able to detect the appearance
of $\nu_\tau$'s from
$\nu_\mu\to\nu_\tau$ oscillations
using the decay mode
$\tau\to\pi\nu_\tau$.
They expect a sensitivity for such oscillations of
$\sin^2 2\theta \gsim 0.21$
for $\Delta m^2 \gsim 2\times 10^{-2}$ eV$^2$.

\begin{figure}[ht]
\psfig{figure=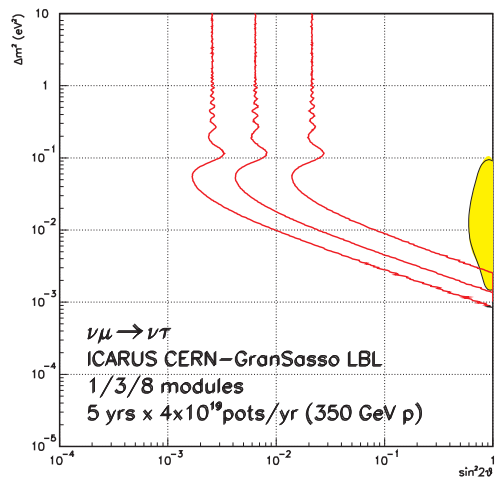,width=2.6in}
\caption[]{ICARUS excluded region 
for $\nu_\mu \to \nu_\tau$ oscillations 
if no signal is observed \cite{ref:Bueno}.
The region favored by Super-K
is shown near $\sin^2 2\theta = 1$,
with $\Delta m^2$ between $10^{-1} - 10^{-3}$ eV$^2$.}
\label{fig:icarustau}
\end{figure}

At CERN, the {\bf N}eutrino Beam to {\bf G}ran {\bf S}asso
(NGS) project plans on building several neutrino detectors
in the Gran Sasso Laboratory in central Italy,
732 km from the CERN wide-band $\nu_\mu$ beam.
The energy and flux of the CERN $\nu_\mu$ beam
are both somewhat higher than is planned for NuMi at Fermilab.
The ICARUS experiment \cite{ref:Bueno}, 
using liquid argon TPC as a target and detector,
is optimized for $\nu_\tau$ and $\nu_e$ appearance.
It is approved and expects to be taking data 
as soon as the beam is available (2003?).
They can search for $\nu_\mu\to \nu_e$ appearance down to
$\sin^2 2\theta \gsim 1\times 10^{-3}$,
and $\nu_\mu\to \nu_\tau$ appearance down to
$\sin^2 2\theta \gsim 5\times 10^{-3}$,
for $\Delta m^2 \gsim 2\times 10^{-2}$ eV$^2$.
A projected exclusion plot is shown in Fig.~\ref{fig:icarustau}.

Several other detectors are being proposed: OPERA \cite{ref:Bueno}
(a lead-emulsion stack),
NICE (iron and scintillator),
AQUA-RICH (water $\check{\mbox{C}}$erenkov), and NOE.
The NOE detector \cite{ref:Scapparone}
will be consist of TRD and calorimeter modules,
optimized for the detection of electrons from
$\nu_\mu\to \nu_e\to e X$, 
and also and from $\nu_\mu\to\nu_\tau\to \tau X$,
$\tau\to e\nu\nu$ (with missing $p_t$).
They also hope to measure the rate for neutral current events
relative to charged current events.

If the interpretation of the Super-K data is correct,
we will soon have a wealth of data to pin down
the oscillation parameters for
$\nu_\mu\lra\nu_e$, $\nu_\tau$, and $\nu_{sterile}$.

\subsection{High luminosity $e^+e^-$ Colliders}
\label{ss-collider}

The IHEP Laboratory in Beijing has been operating
the BEPC collider and the BES detector,
with $e^+e^-$ collisions at or above tau pair threshold,
for many years. They are pursuing the physics of taus 
(their measurement of the tau mass totally dominates
the world average), $\psi$ spectroscopy and decay,
and charm.
They have proposed the construction
of a much higher luminosity tau-charm factory (BTCF).
However, 
because of the limitation of funds, the BTCF will 
not be started for at least 5 years \cite{ref:Qi}.
The plan for the near term is to continue to improve
the luminosity of the existing BEPC collider.
In tau physics, they
hope to produce results from the BES experiment at BEPC
on $m_{\nu_\tau}$, where they favor \cite{ref:Qi}
the use of the decay mode $\tau\to K\bar{K}\pi\nu_\tau$.
  
Within the next two years, three new ``B Factories'',
high luminosity $e^+e^-$ colliders with
center of mass energies around 
10.58 GeV (the $\Upsilon(4S) \to B\bar{B}$ resonance) 
will come on line:
the upgraded CLEO III detector at the CESR collider;
BaBar at SLAC's PEP-II collider \cite{ref:Seiden};
and BELLE at KEK's TRISTAN collider \cite{ref:Oshima}.
All these colliders and detectors expect to begin operation
in 1999. The BaBar and BELLE experiments operate with asymmetric beam
energies, to optimize the observation of CP violation in the $B$ 
mixing and decay; CLEO will operate with symmetric beams.

At design luminosities, these experiments expect to collect
between $10^7$ and $10^8$ tau pair events per year.
In a few years, one can expect more than $\sim 10^{8}$ events
from BaBar, BELLE, and CLEO-III.
The asymmetric beams at BaBar and BELLE should not present
too much of a problem (or advantage) for tau physics;
it may help for tau lifetime measurements.
The excellent $\pi$-$K$ separation that the detectors
require for their $B$ physics goals will also
be of tremendous benefit in the study of
tau decays to kaons.
BaBar and BELLE will also have improved ability
to identify low-momentum muons,
which is important for Michel parameter measurements.

The high luminosities will make it possible
to improve the precision of almost all the 
measurements made to date in tau decays,
including the branching fractions, Michel parameters,
and resonant substructure in multi-hadronic decays.
In particular, they will be able to 
study rare decays (such as $\eta X\nu_\tau$ or $7\pi\nu_\tau$)
with high statistics.
They will search with higher sensitivity for
forbidden processes, such as LFV neutrinoless decays
and CP violation due to an 
electric dipole moment \cite{ref:Seiden,ref:Oshima}
in tau production or a charged Higgs in tau decay.
And they  will be able to search for neutrino masses
down to masses below the current 18 MeV/c$^2$ best limit.

At the next Tau Workshop, we look forward to a raft of new results
from DONUT, BES, the B Factories, the Tevatron, and the long-baseline
neutrino oscillation experiments.
I expect that that meeting will be full of 
beautiful results, and maybe some surprises.

\section{Acknowledgements}
\label{s-ack}

I thank the conference organizers for a thoroughly stimulating 
and pleasant conference.
This work is supported by the U.S.~Department of Energy.

\end{document}